\newcommand{\AP}[3]{Ann.\ Phys.\ {\bf #1},\ #2 (#3)}
\newcommand{\NPA}[3]{Nucl.\ Phys.\ {\bf A#1},\ #2 (#3)}
\newcommand{\NPB}[3]{Nucl.\ Phys.\ {\bf B#1},\ #2 (#3)}

\newcommand{\PLB}[3]{Phys.\ Lett.\ B\ {\bf #1},\ #2 (#3)}
\newcommand{\PR}[3]{Phys.\ Rep.\ {\bf #1},\ #2 (#3)}
\newcommand{\PRL}[3]{Phys.\ Rev.\ Lett.\ {\bf #1},\ #2 (#3)}
\newcommand{\PRA}[3]{Phys.\ Rev.\ A\ {\bf #1},\ #2 (#3)}

\newcommand{\PRC}[3]{Phys.\ Rev.\ C\ {\bf #1},\ #2 (#3)}
\newcommand{\PRD}[3]{Phys.\ Rev.\ D\ {\bf #1},\ #2 (#3)}
\newcommand{\JPG}[3]{J.\ Phys.\ G\ {\bf #1},\ #2 (#3)}

\newcommand{\ZPC}[3]{Z.\ Phys.\ C\ {\bf #1},\ #2 (#3)}

\newcommand{\EPJA}[3]{Eur.\ Phys.\ J.\ A\ {\bf #1},\ #2 (#3)}
\newcommand{\PTP}[3]{Prog.\ Theo.\ Phys.\ {\bf #1},\ #2 (#3)}

\newcommand{\diracslash}[1]{#1\llap{/\kern2pt}}

\newcommand{\be}{\begin{equation}}
\newcommand{\ee}{\end{equation}}
\newcommand{\bea}{\begin{eqnarray}}
\newcommand{\eea}{\end{eqnarray}}
\newcommand{\ba}[1]{\begin{array}{#1}}
\newcommand{\ea}{\end{array}}

\documentclass[twocolumn,prd,showpacs,preprintnumbers,amsmath,amssymb,aps,float,floatfix]{revtex4}
\usepackage{graphicx}
\usepackage{dcolumn}
\usepackage{bm}
\begin{document}
\preprint{}
\title{Bulk viscosity in  hyperonic star and r-mode instability}
\author{T. K. Jha$^{1,2}${\footnote {email:tkjha@bits-goa.ac.in}},
H. Mishra$^1${\footnote {email: hm@prl.res.in}} and
V. Sreekanth$^1${\footnote {email: skv@prl.res.in}}}
\affiliation {${}^1$  Theoretical Physics Division, Physical Research Laboratory, Navrangpura, Ahmedabad - 380 009, India \\
 ${}^2$ Physics Group, BITS Pilani Goa Campus, Goa - 403 726, India}
\date{\today}
\def\be{\begin{equation}}
\def\ee{\end{equation}}
\def\bearr{\begin{eqnarray}}
\def\eearr{\end{eqnarray}}
\def\zbf#1{{\bf {#1}}}
\def\bfm#1{\mbox{\boldmath $#1$}}
\def\hf{\frac{1}{2}}
\begin{abstract}
We consider a rotating neutron star with the presence of hyperons in 
its core, using an equation of state in an effective chiral model 
within the relativistic mean field approximation. We calculate the 
hyperonic bulk viscosity coefficient due to nonleptonic weak interactions. 
By estimating the damping timescales of the dissipative processes,
we investigate its role in the suppression of gravitationally driven 
instabilities in the $r$-mode. We observe that $r$-mode instability 
remains very much significant for hyperon core temperature of around 
$10^8 ~$K, resulting in a comparatively larger instability window. 
We find that such instability can reduce the angular velocity of the 
rapidly rotating star considerably upto $\sim0.04~\Omega_K$, 
with $\Omega_K$ as the Keplerian angular velocity. 
\end{abstract}
 \maketitle

\section{INTRODUCTION}
The study of neutron stars are natural testing ground for studying 
extremely dense matter. The densities in the interior of such stars 
can reach up to several times the nuclear matter saturation density.
At such high densities, with higher fermi momenta being available, 
high mass hadrons can be accommodated leading to a hyperonic core 
in the neutron star interior \cite{hypstar,weber}. There could also 
be the possibility that at such densities, when the nucleons are 
crushed, there could be quark matter \cite{qmatter} which can result 
in a color superconducting core in the interior of the neutron star
\cite{colscrev}. 
In fact, there are different possibilities of the ground state of dense
matter, which could be stable strange quark matter \cite{witten},
and various possibilities of color superconducting matter \cite{colsc}. 
The reason is that the true ground state of the
dense matter system for densities relevant for the
densities in the interior of neutron stars is still an open problem 
because of the inherent nonperturbative nature of strong interaction physics.
Moreover, external conditions like  electrical and color charge 
neutrality conditions for the bulk matter in the interior of the star
can also lead to various different possible phases of quark matter
\cite{chargen,hmamchargen}.
This has given rise to various possibilities of 
compact stellar objects like
neutron stars, strange stars, hyperonic stars or hybrid stars with a 
quark matter core and a crust of hadronic matter \cite{weber}.
Obviously, it  becomes
very challenging to distinguish various compact stellar objects observationally.

One of the various signatures suggested to distinguish different compact stars
has been the r-mode instability \cite{rmoderev,jai,rischkehuang}.
Various pulsating modes exist in neutron stars
classified by the nature of the restoring forces. The r-modes correspond to the
pulsating modes of the rotating stars where the restoring force is the Coriolis
force and these modes are axial modes \cite{andersonn}. As the fluid inside 
the star is self 
gravitating, these oscillations can couple to metric perturbations and lead to
emission of gravitational waves \cite{lindblom}. Gravitational modes drive 
the r-modes
unstable due to Chandrasekhar-Friedman- Schutz (CFS) mechanism \cite{cfs}. This
phenomenon will occur if the damping is sufficiently small and therefore 
provides a probe to study the viscosity of the matter in the interior 
of the star.
While shear viscosity prevents differential rotation in a star, bulk viscosity
dampens the density fluctuations in the star. Shear viscosity seems to 
be important at lower temperatures while bulk viscosity becomes the dominating
dissipation mechanism at higher temperatures ($\sim  10^9 K$). Further,
since typical r-mode frequencies are of the order of rotational frequencies 
of the stars ($ 1s^{-1}<\omega<10^3 s^{-1}$), which are of the order of 
weak interactions, therefore the viscosity that is of relevance for
r-mode instability will be dominated by the weak processes. 
Bulk viscosity is produced when the mode oscillations induce perturbations in
pressure and density and drives the system away from $\beta$-equilibrium.
As a result energy is dissipated from the system as the 
weak interaction tries to
restore the equilibrium. While for hadronic neutron star matter,
modified Urca processes ($n+n\rightarrow n+p+e^-+\bar\nu_e$) 
involving leptons are important,
it has been noted that  the damping of the instability is dominated by
large bulk viscosity arising due to nonleptonic processes involving hyperons
\cite{deba}.
The corresponding viscosities are not only stronger, the temperature 
dependence is also different (varying as $T^{-2}$ instead of $T^6$ ) which 
makes the hyperon bulk viscosity important at lower temperatures. 
Bulk viscosities of dense baryonic matter have been calculated under various
 assumptions as well as conditions over last few years
\cite{jones,drago,deba,LO,mohit,lom98,dieperink}.

The study of the r-mode oscillations provides an avenue to study the
density profile of the star along with the mass and the radius which
in turn depends upon the equation of state for the matter in the 
interior of the star. Phenomenologically, for hadronic matter
parallel to the $\sigma-\omega$ model, popularly known as
Walecka model \cite{walecka,serot}, the chiral effective models have been
developed and and are applied to nuclear matter. A chiral $\sigma-\omega$
model along with dilatons in the context of phase transition has been
developed in Ref.\cite{juergen} and further generalised to
describe strange hadronic matter\cite{juergenn,verena,amruta}. 
Another approach that 
was also considered is the parity doublet model \cite{juergennn}.
We had recently considered attributes of
rotating hyperonic star within a chiral $\sigma$ 
model \cite{prakash87, glendenningnpa,dutta}.
This hadronic model also embodies a dynamical
generation of the vector meson mass 
along with nonlinear terms
in the scalar field interactions to reproduce nuclear saturation properties
at reasonable incompressibility \cite{pksahu,pksahuu}. This model
was then generalised to include the lowest lying octet
of baryons \cite{tkj06,tkj08,tkj09}.
The resulting equation of state was used to calculate
various gross properties of nonrotating as well as rotating neutron stars.
It was further observed that the parameters of the model are sternly 
constrained and turn to be consistent with the constraints on nuclear 
equation of state from heavy ion collision data \cite{tkj08,tkj09}. 
In the present work, we discuss the calculation of bulk viscosity 
within this model and its consequences regarding the r-mode instability 
problem in rotating neutron star with a hyperonic core. 

We organise the paper as follows. In the first subsection of section II, 
we describe very briefly the chiral model that we shall be using to 
derive the equation of state needed to study the structure of the 
neutron star. In the subsequent subsections in the section, we derive 
the bulk viscosity of the hyperonic matter and discuss their role 
in r-mode damping. In section III, we discuss the results of the 
present calculations. Finally, in section IV, we summarize the 
results and give possible outlook of the present investigation.

\section{FORMALISM}
\subsection{Effective chiral model and equation of State}\label{eos}
In this subsection, we briefly summarise the salient features
of the chiral model including hyperons, which has been considered earlier
in the context of neutron star matter. It will be used later to
calculate the bulk viscosity. 

The  effective chiral model that we shall consider is a generalisation 
of the model considered in Ref.\cite{tkj06} 
to include the lowest lying octet of baryons 
($n,p,\Lambda^{0},\Sigma^{-,0,+},\Xi^{-,0}$). They interact
via the exchange of the pseudo-scalar mesons $\pi$, the scalar 
meson $\sigma$, the vector meson $\omega$ and the iso-vector $\rho-$meson. 
Lagrangian density under consideration is 
given by \cite{tkj06,tkj08,tkj09}:

\begin{eqnarray}
\label{lag}
{\cal L}&=& \bar\psi_B~ \big(i\gamma_\mu\partial^\mu
         - g_{\omega B}\gamma_\mu\omega^\mu
         - \frac{1}{2}g_{\rho B}{\vec \rho}_\mu\cdot{\vec \tau}
            \gamma^\mu\big )~ \psi_B
\nonumber \\
&& 
      - g_{\sigma B~}~\bar\psi_B~\big(\sigma + i\gamma_5
             \vec \tau\cdot\vec \pi \big)~ \psi_B
\nonumber \\
&&
        + \frac{1}{2}\big(\partial_\mu\vec \pi\cdot\partial^\mu\vec\pi
        + \partial_{\mu} \sigma \partial^{\mu} \sigma\big)
\nonumber \\
&&
        - \frac{\lambda}{4}\big(x^2 - x^2_0\big)^2
        - \frac{\lambda B}{6}\big(x^2 - x^2_0\big)^3
        - \frac{\lambda C}{8}\big(x^2 - x^2_0\big)^4
\nonumber \\
&&      - \frac{1}{4} F_{\mu\nu} F_{\mu\nu}
        + \frac{1}{2}{g_{\omega B}}^{2}x^2 \omega_{\mu}\omega^{\mu}
\nonumber \\
&&      - \frac {1}{4}{\vec R}_{\mu\nu}\cdot{\vec R}^{\mu\nu}
        + \frac{1}{2}m^2_{\rho}{\vec \rho}_{\mu}\cdot{\vec \rho}^{\mu}\ .
\end{eqnarray}

The first two lines of the above Lagrangian density represent the interaction 
of baryons $\Psi_B$
with the aforesaid mesons. In the next two lines, we have the kinetic and 
the non-linear terms in the pseudoscalar-isovector pion field `$\zbf \pi$',  
the scalar field `$\sigma$', and with $x^2= {\zbf \pi}^2+\sigma^{2}$.
Finally in the last two lines, we have the field strength and the mass 
term for the vector field `$\omega$' and the iso-vector field 
`$\vec \rho$' meson. The terms in equation (1) with the subscript $'B'$ 
should be interpreted as sum over the states of lowest baryonic 
octet. In this paper, we shall be concerned only with the normal 
non-pion condensed state of matter and, so we take $<\zbf \pi>=0$.

The interaction of the scalar and the pseudoscalar mesons with the vector 
boson generates a dynamical mass for the vector bosons through 
spontaneous breaking of the chiral symmetry with scalar field getting the 
vacuum expectation value $x_0$. Then the masses of the baryons, 
the scalar and the vector mesons, are respectively given by

\begin{eqnarray}
m_B = g_{\sigma B} x_0,~~ m_{\sigma} = \sqrt{2\lambda} x_0,~~
m_{\omega} = g_{\omega B} x_0\ .
\end{eqnarray}

In the above, $x_0$ is the vacuum expectation value of the $\sigma$ field.
We could have taken an interaction of the $\rho-$meson with the scalar
and the pseudoscalar mesons similar to the omega meson. However, a
dynamical mass
generation mechanism of the $\rho-$meson in a similar manner will not generate 
the correct symmetry energy. Therefore, we have taken an explicit mass term
for the isovector $\rho-$meson similar to what was considered in earlier 
works \cite{tkj06,tkj08,dutta}.
 
We use mean-field approximation (\cite{walecka}) to evaluate the meson fields 
in our present
calculations. In the mean-field treatment, one assumes the mesonic fields to be 
uniform, i.e., without any quantum fluctuations. 
The details of the model that we use in our present investigation and 
its attributes such as the derivation of the equation of 
motion of the meson fields and its equation of state $(\varepsilon~\&~P)$
of the many baryonic system, can be found in our preceding work \cite{tkj08}.
The vector meson and the iso-vector meson field equations in the mean field 
framework are then given by

\begin{eqnarray}
\omega_0=\sum_{B}\frac{ \rho_B }{g_{\omega B} x^2} \ , \label{omega}
\end{eqnarray}

\begin{eqnarray}
\rho_{03} =\sum_{B} \frac{g_{\rho B}}{m_\rho^2} I_{3 B}\rho_{B}\ . \label{rho}
\end{eqnarray}

In the above equations the quantity $\rho_B$ is the baryon density 
and $I_{3 B}$ is the third component of the isospin of each baryon species.

The scalar field equation can be written in terms of the variable
$Y=x/x_0$ with $x=(<\sigma^2+\pi^2>)^{1/2}$ as \cite{tkj06,tkj08}

\begin{eqnarray}
\sum_B &&[(1-Y^2) -\frac{B}{c_{\omega B}}(1-Y^2)^2
+\frac{C}{c_{\omega B}^2}(1-Y^2)^3 
\nonumber \\
&&
+\frac{2 c_{\sigma B}c_{\omega B}\rho_B^2}{m_{B}^2Y^4}
-\frac{2 c_{\sigma B}\rho_{SB}}{m_{B} Y}]=0 ,
\label{effmass}
\end{eqnarray}
\noindent
where the effective mass of the baryonic species is $m_{B}^{\star} 
\equiv Ym_{B}$ and $c_{\sigma B}\equiv  g_{\sigma B}^2/m_{\sigma}^2 $ are the
$c_{\omega B} \equiv g_{\omega B}^2/m_{\omega}^2 $ are the usual
scalar and vector coupling constants respectively. 
Similarly, in the present model describing dense matter, the $\omega-$meson 
mass is generated dynamically. This vector meson mass enters in Eq. 
(\ref{effmass}) through the ratio $c_{\omega} = 
(g_{\omega}/m_{\omega})^2 \equiv 1/{x_0}^2$.
Various parameters of the model for the nuclear matter case
(i-th meson nucleon couplings, $C_\sigma,C_\omega, C_\rho$ and the nonlinear 
couplings B and C)
are fitted from nuclear matter saturation properties
\cite{dutta,tkj04,tkj06}. It was shown recently that these parameters
are rather constrained by the nuclear matter saturation properties
like the binding energy per nucleon, saturation density, the nuclear
incompressibility, as well as the  asymmetry energy, the
effective mass of the nucleon
and the pion decay constant $f_\pi$.
Further in Eq.(\ref{effmass}), the quantities $\rho_B$ and 
$\rho_{SB}$ are the baryon density and the scalar density for a given
baryon species given respectively as,

\begin{equation}
\rho_B= \frac{\gamma}{(2\pi)^3}\int^{k_B}_o d^3k,
\end{equation}

\begin{equation}
\rho_{SB}= \frac{\gamma}{(2\pi)^3}\int^{k_B}_o\frac{m^*_{B} d^3k}
         {\sqrt{k^2+m_{B}^{\star 2}}},
\end{equation}
\noindent
where $k_B$ is the fermi momentum of the baryon and $\gamma=2$ is the spin
degeneracy factor.

For a given baryon density, the total energy density `$\varepsilon$' 
and the pressure `$P$' can be written in terms of the dimensionless 
variable $Y=x/x_0$ as

\begin{eqnarray}
\label{ep0}
\varepsilon
&=&
 \frac{2}{\pi^2}\int^{k_B}_0 k^{2}dk{\sqrt{k^2+m_B^{\star 2}}}
         +  \frac{m_B^2(1-Y^2)^2}{8c_{\sigma B}}
\nonumber \\
&&
        - \frac{m_B^2 B}{12c_{\omega B}c_{\sigma B}}(1-Y^2)^3
        + \frac{m_B^2 C}{16c_{\omega B}^2c_{\sigma B}}(1-Y^2)^4
\nonumber \\
&&
        + \frac{1}{2Y^2}{c_{\omega_{B}} \rho_B^2}
        +\frac{1}{2}m_{\rho}^{2}\rho_{03}^2
\nonumber \\
&&
        + \frac{1}{\pi^{2}}\sum_{\lambda=e,\mu^{-}}\int^
           {k_\lambda}_0 k^{2}dk{\sqrt{k^2+m^2_{\lambda}}}\ ,
\end{eqnarray}
\begin{eqnarray}
P &=&
          \frac{2}{3\pi^2}\int^{k_B}_0\frac{k^{4}dk}
          {{\sqrt{k^2+m_B^{\star 2}}}}
        - \frac{m_B^2(1-Y^2)^2}{8c_{\sigma B}}
\nonumber \\
&&
        + \frac{m_B^2 B}{12c_{\omega B}c_{\sigma B}}(1-Y^2)^3
        - \frac{m_B^2 C}{16c_{\omega B}^2c_{\sigma B}}(1-Y^2)^4
\nonumber \\
&&
        + \frac{1}{2Y^2}{c_{\omega_{B}} \rho_B^2}
        + \frac{1}{2}m_{\rho}^{2}\rho_{03}^2\
\nonumber \\
&&
       +\frac{1}{3\pi^2}\sum_{\lambda=e,\mu^{-}}\int^{k_\lambda}_0
          \frac{k^{4}dk}{{\sqrt{k^2+m^2_{\lambda}}}}
\end{eqnarray}

The terms in eqns. (3) and (4) with the subscript $`B'$ should be interpreted 
as sum over all the states of the baryonic octets. The meson field equations 
for the $\sigma$, $\omega$ and $\rho-$mesons are then solved self-consistently 
at a fixed baryon density to obtain the respective field strengths.
The EoS for the $\beta-$equilibrated for the hyperon rich matter is obtained
with the requirements of conservation of total baryon number and charge
neutrality condition given by \cite{tkj06}

\begin{equation}
\sum_{B}Q_{B}\rho_{B}+\sum_{l}Q_{l}\rho_{l}=0,
\end{equation} 
\noindent
where $\rho_{B}$ and $\rho_{l}$ are the baryon and the lepton (e,$\mu$) number
densities with $Q_{B}$ and $Q_{l}$ as their respective electric charges.

\subsection{Hyperon Bulk-viscosity}
Bulk viscosity characterizes the response of the system to an 
externally oscillating change in the volume. The volume expansion or
contraction leads to a change in density or the chemical potential 
of the system. This drives the system out of chemical equilibrium. 
The equilibrium is restored by the microscopic processes. If the 
equilibrium time scales are comparable to the oscillating time scales, 
there will be energy dissipation. In the context of neutron star, the 
typical oscillation frequencies are less than a kilo hertz. Therefore, 
the microscopic processes that will be relevant are the weak processes.

It is already known that the non-leptonic processes containing 
hyperons lead to high values of bulk viscosity for rotating neutron 
stars with temperature ~ $10^9-10^{10}$K \cite{jones,LO,pbjones}. 
The leptonic processes are suppressed by smaller phase space factors.
Thus the  relevant reactions which are going to give a lower limit 
on the rates (or upper limit on bulk viscosity) are
\begin{eqnarray}
 n+n &\longleftrightarrow& p+\Sigma^- \label{r1} \\
 n+p &\longleftrightarrow& p+\Lambda^0\label{r2} \\
 n+n &\longleftrightarrow& n+\Lambda^0\label{r3}
\end{eqnarray}

The coefficient of bulk viscosity relates difference between
the perturbed pressure $p$ and the thermodynamic pressure $\tilde p$
to the macroscopic expansion of the fluid as 

\begin{equation}
p-\tilde {p} = -\zeta \zbf{\nabla}\cdot{\zbf v}
\end{equation}
where $\zbf v$ is the velocity of the fluid element and $\zeta$ is the 
coefficient of bulk viscosity, which, in general, is complex in nature 
\cite{Landau}.

The relativistic expression for the real part of $\zeta$, which 
corresponds to the damping, can be calculated in terms of microscopic 
equilibrium restoring reaction rates \cite{LO, Landau}. Within a
relaxation time approximation, the real part of $\zeta$ is given as,
\begin{equation}
\zeta = \frac{p\left(\gamma_{\infty}-\gamma_{0}\right) \tau}{1+\left( \omega\tau\right)^2 } \label{zeta}
\end{equation}
where $\gamma_{\infty}$ and $\gamma_0$ are the ``infinite'' and 
``zero'' frequency adiabatic index respectively. $\omega$ is the 
angular frequency of the perturbation in co-rotating frame 
and $\tau$ is the net equilibrium restoring microscopic relaxation 
time. The expression for $\gamma_{\infty}-\gamma_{0}$ is
\begin{equation}
\gamma_{\infty}-\gamma_{0} = -\frac{{n_B}^2}{p}
\frac{\partial p}{\partial n_n}\frac{d\tilde{x}}{dn_B} 
\label{gammas}
\end{equation}
Here $n_B$ corresponds to the total baryon density and 
$\tilde{x} = \frac{n_n}{n_B}$ is the neutron fraction.Thus the
difference $\gamma_\infty-\gamma_0$  can be calculated from a given
equation of state. The angular frequency 
$\omega$ of the $r$ - mode ($l$=2, $m$=2) in a co-rotating frame is given 
in terms of the Keplerian frequency $\Omega_K$ 
of the rotating star as $\omega= \frac{2m}{l(l+1)}\Omega_K$ (\cite{NAnd}).

The prominant reactions involving the lightest hyperons, $\Sigma^-$ 
and $\Lambda^0$, which have higher population in a given star 
are as given by equations (\ref{r1}) and (\ref{r2}). 
The rates of these reactions can be calculated from the tree-level
Feynman diagrams involving the exchange of a W boson.

We are not be considering the other reaction 
$n+n \longleftrightarrow n+\Lambda^0$ since it has no simple
W-boson exchange picture. We shall discuss more regarding this
in Section III. The relaxation time $\tau$ (at a temperature $T$), 
when both $\Sigma^{-}$ and $\Lambda^0$ are present, is given 
by \cite{LO,mohit}
\begin{equation}
 \frac{1}{\tau}=\frac{(k_BT)^2}{192{\pi^3}}
\left(k_{\Sigma}\left\langle\vert\mathcal M_{\Sigma}^2\vert  \right\rangle
 + k_{\Lambda}\left\langle\vert\mathcal M_{\Lambda}^2\vert  \right\rangle\right) \frac{\delta \mu}{n_B \delta x_{n}} \label{tau}
\end{equation}
where $k_B$ is the Boltzmann's constant and $k_{\Lambda}$ and $k_{\Sigma}$ 
are the Fermi momenta of these hyperons.
 $\delta \mu\equiv\delta \mu_n - \delta \mu_{\Lambda}$ is the chemical 
potential imbalance. 
$\delta x_{n}=x_n-\tilde{x_n}$ is the small difference between the 
perturbed and equilibrium values of the neutron fraction. 
$\left\langle\vert\mathcal M^2\vert  \right\rangle$ are the angle averaged, 
squared, summed over initial spinors matrix elements of the 
reactions calculated from the Feynman diagrams. We refer 
\cite{LO,mohit} for the expressions for 
$\left\langle\vert\mathcal M_{\Sigma}^2\vert  \right\rangle$ 
and $\left\langle\vert\mathcal M_{\Lambda}^2\vert  \right\rangle$. 
We note that the contribution from $\Lambda$ hyperons will not be 
present in Equation (\ref{tau}) while considering a  neutron star medium 
before the appearance of $\Lambda$.

The factor $\delta \mu / n_B \delta x_{n}$ is determined from the 
constraints imposed by the electric charge neutrality and the baryon number 
conservation given respectively as
\begin{eqnarray}
\delta x_p - \delta x_\Sigma &=& 0\\
\delta x_n + \delta x_\Lambda +\delta x_p +\delta x_\Sigma &=&0
\end{eqnarray}
together with the condition that the non-leptonic strong interaction reaction
\begin{equation}
 n+\Lambda^0 \longleftrightarrow p+\Sigma^- \label{r4}
\end{equation}
which has a higher rate, is in equilibrium compared to weak interaction
processes giving rise to the bulk viscosity. Equilibrium of 
this reaction implies that both the reactions (\ref{r1}) and 
(\ref{r2}) have equal chemical potential imbalance,
\begin{equation}
\delta \mu\equiv\delta \mu_n - \delta \mu_{\Lambda} = 2\delta \mu_n -\delta \mu_p - \delta \mu_{\Sigma}.
\end{equation}
Using these constraints, we can write,
\begin{eqnarray}\label{delmu1}
\frac{\delta \mu}{n_B \delta x_{n}} &=& \alpha_{nn} + 
\frac{(\beta_n-\beta_{\Lambda})(\alpha_{np}-\alpha_{\Lambda p}
+\alpha_{n\Sigma}-\alpha_{\Lambda\Sigma})}{2\beta_{\Lambda}
-\beta_{p}-\beta_{\Sigma}}
\nonumber \\
&& -\alpha_{\Lambda n} - \frac{(2\beta_n-\beta_p-\beta_{\Sigma})
(\alpha_{n\Lambda}-\alpha_{\Lambda\Lambda})}{2\beta_{\Lambda}
-\beta_{p}-\beta_{\Sigma}}
\end{eqnarray}
where $\alpha_{ij}=\left(\frac{\partial\mu_i}{\partial n_j} \right)_{n_{k},k\neq j}$ and $\beta_i = \alpha_{ni} + \alpha_{\Lambda i}-\alpha_{pi}-\alpha_{\Sigma i}$. 
These expressions are for the case when both the $\Sigma^-$ and 
$\Lambda^0$ hyperons are present. One can not use the reaction (\ref{r4}) 
while considering the region where we have only $\Sigma^-$ hyperons. 
In that case, instead of Eq.(\ref{delmu1}), we have
\begin{eqnarray}\label{delmu2}
\frac{2\delta \mu}{n_B \delta x_{n}} &=& 4\alpha_{nn}
 -2(\alpha_{pn}+\alpha_{\Sigma n}+\alpha_{np}+\alpha_{n\Sigma})
\nonumber \\
&& +\alpha_{pp}+\alpha_{\Sigma p}+\alpha_{p\Sigma}+\alpha_{\Sigma \Sigma}.
\end{eqnarray}
Now the $\alpha_{ij}$'s can be found out using the expression 
for baryon chemical potential and the equations of motion of the 
mesonic fields given by equations (\ref{omega},\ref{rho}).
In general, the form of the $\alpha_{ij}$ is given as,
\begin{eqnarray}\label{alpha1}
  \alpha_{ij}&=&\frac{m^*_i m_i}{\sqrt{k_{F_{i}}^2+m_{i}^{*2}}}
\frac{\partial Y}{\partial n_{j}} 
+ \frac{g_{\omega i}}{g_{\omega j}}\frac{1}{(Yx_{0})^2}\\
&&
- \frac{2 g_{\omega i}x_{0}}{(Yx_{0})^3} \left(\sum_{B}
\frac{ \rho_B }{g_{\omega B}} \right) \frac{\partial Y}{\partial n_{j}}
+\frac{1}{2 m_{\rho}^2}(g_{\rho i}g_{\rho j})(I_{3i}I_{3j})\nonumber 
\end{eqnarray}
for $i\neq j$ and 
\begin{eqnarray}\label{alpha2}
  \alpha_{ii}&=&\frac{m^*_i m_i}{\sqrt{k_{F_{i}}^2+m_{i}^{*2}}}\frac{\partial Y}{\partial n_{i}} + \frac{\pi^2}{k_{F_{i}}\sqrt{k_{F_{i}}^2+m_{i}^{*2}}}
+ \frac{1}{(Yx_{0})^2}
\nonumber \\
&&
- \frac{2 g_{\omega i}x_{0}}{(Yx_{0})^3} \left(\sum_{B}\frac{ \rho_B }{g_{\omega B}} \right) \frac{\partial Y}{\partial n_{i}}
+\frac{1}{2}(\frac{g_{\rho i}I_{3i}}{m_{\rho}})^2 .
\end{eqnarray}
As before, $Y$ is the scalar field expectation value in the medium
in units of its vacuum expectation value. Further,
$\frac{\partial Y}{\partial n_{i}}$ are calculated from the scalar
field equation Eq.(\ref{effmass}) and are given by
\begin{equation}\label{dY}
\frac{\partial Y}{\partial n_{i}} = \frac{1}{D}\left(\frac{2 c_{\sigma i}c_{\omega i}\rho_i}{m_{B}^2Y^4}
-\frac{c_{\sigma i}m^*_i}{m_i Y \sqrt{k_{F_{i}}^2+m_{i}^{*2}}}\right)
\end{equation}
with 
\begin{eqnarray}\label{D}
D&=&\sum_B[ Y +\frac{2B}{c_{\omega B}}(Y^2-1) Y +
\frac{3C}{c_{\omega B}^2}(Y^2-1)^2 Y 
\nonumber \\
&&
-\frac{c_{\sigma B}\rho_{SB}}{m_B Y^2}+
\frac{4 c_{\sigma B}c_{\omega B}\rho_B^2}{m_{B}^2Y^5}\\
&&
+\frac{c_{\sigma B}}{m_B Y} \frac{\gamma}{(2\pi)^3}
\int^{k_{F_{B}}}_o d^3k \frac{k^2 m_B}{(k^2+m^{*2}_B)^{3/2}}] 
\nonumber 
\end{eqnarray}
Using equations (\ref{delmu2}-\ref{D}), one can compute the relaxation 
time from Eq.(\ref{tau}) and hence the bulk viscosity given 
in Eq.(\ref{zeta}), for a given equation of state.

\subsection{R-mode damping}

As mentioned in the introduction section, the r-modes correspond to 
the axial modes where the restoring force is the Coriolis force. 
The r-mode frequency $\omega_c$ in the corotating frame, to first 
order in $\Omega$, the rotation frequency of the star  is  given by 
\cite{provost}
\be
\omega_c=\frac{2m\Omega}{l(l+1)}+O(\Omega^3)
\ee
The r-modes correspond to $l=m$\cite{papaloizou}.
The r-mode frequency observed by an inertial observer is given by
\be
\omega_0=\omega_c-m\Omega=\left(\frac{2}{l(l+1)}-1\right)m\Omega.
\ee
Thus, $l=m=2$ mode becomes most unstable to emission of gravitational waves
due to Chandrasekhar Friedman Shultz  (CFS) mechanism.
In a rotating star emission of these waves causes the modes to grow. 
This instability can get damped due to the various viscosities
of the stellar matter. This happens when the damping time scales associated 
with these viscous processes are comparable to the gravitational radiation
(GR) time scale.

We need the expressions for the time scales associated with the
dissipative processes and GR in order to understand the nature of 
the damping of the r-modes. The imaginary part of the dissipative 
time scale (which causes the damping) is given by \cite{LO,basil}
\begin{equation}
 \frac{1}{\tau_i} = -\frac{1}{2\tilde E}\left(\frac{d\tilde E}{dt}\right)_i \label{taui}
\end{equation}
where $i$ labels the various dissipative phenomena like hyperonic bulk 
viscosity (B), bulk viscosity due to Urca processes (U), 
shear viscosity ($\eta$) GR etc. 
In the above, $\tilde E$ is  energy of the r-mode in the co-rotating frame.
This can arise both from velocity perturbation as well as the perturbation 
of the gravitational potential. For a slowly rotating star, the dominant 
contribution is from the velocity perturbation and is given as
\begin{equation}
\tilde E = \frac{1}{2}\int \rho |\delta \vec v|^2 d^{\,3}x,\label{tildeE}
\end{equation}
with, $\rho(r)$ being the mass density of the star.
Assuming the spherical symmetry, mode energy can be reduced into an
one dimensional integral \cite{lom98} as
\begin{equation}
\tilde{E}=\frac{1}{2}\alpha^2 \Omega^2 R^{-2l+2}\int_0^R \rho(r) r^{2l+2} dr.
\label{mode enrgy}
\end{equation}
with $l=m=2$ for the r-modes and $R$ denotes the radius
of the star. $\alpha$ is the dimensionless 
amplitude coefficient of the mode, which gets cancelled out in 
the $\tau$ calculation.
This energy is dissipated both by gravitational radiation as well as
thermodynamic transport of the fluid \cite{lindblom,andersonn}. 
The dissipation rate due to the bulk viscosity effects is given by
\begin{equation}
\frac{d\tilde E_B}{dt}=-\int {\rm Re}\,\,\zeta\,\,\,
 |\vec\nabla\cdot\delta\vec v|^2
d^{\,3}x.\label{dedt}
\end{equation}
Here, in general, $|\vec\nabla\cdot\delta\vec v|^2$ depends upon the radial 
and the angular co-ordinates. In slowly rotating stars,
to the lowest order, $\zeta$ depends only on the radius. Therefore, to 
the lowest order in $\Omega$, it is possible to write the bulk viscosity
dissipation rate in Eq.(\ref{dedt}) as an one dimensional integral
by defining a quantity which is the angle averaged expansion squared 
$\left< \vert \vec{\nabla} \cdot \delta \vec{v} \vert ^2 \right>$. 
In terms of this quantity, Eq.(\ref{dedt}) can be written as

\begin{equation}
\frac{d\tilde{E_B}}{dt}=- 4 \pi \int_0^R{\rm Re}\,\,\zeta(r) \left< \vert \vec{\nabla} \cdot \delta \vec{v} \vert^2 \right> r^2 dr.\label{dedtb}
\end{equation}
where $\left< \vert \vec{\nabla} \cdot \delta \vec{v} \vert ^2 \right>$ 
can be determined numerically \cite{lmo99}. However Ref.s\cite{LO,mohit} give 
an analytic expression 
\begin{equation}
\langle|\vec{\nabla} \cdot \delta\vec{v}|^2\rangle = \frac{\alpha^2 
{\Omega}^2} 
{690}
\left(\frac{r}{R}\right)^6\left[1+0.86\left(\frac{r}{R}\right)^2\right]
\left( \Omega^2 \over \pi G\bar\rho \right)^2,
\label{delv}
\end{equation}
which fits to the numerical data. Here $G$ is the gravitational constant 
and $\bar\rho$ is the mean density of the non-rotating star.

Now with the knowledge of density profile $\rho(r)$ of the star, 
it is straightforward to find out the bulk viscosity damping time scales 
from Equations Eq.(\ref{delv}), Eq.(\ref{dedtb}), Eq.(\ref{mode enrgy})
and Eq.(\ref{taui}), once we know the bulk viscosity coefficient $\zeta(r)$.
In the case of bulk viscosity time scale arising due to hyperons 
($\tau_B$), we can get $\zeta(r)$ from Eq.(\ref{zeta}) . 
Similarly we can find out the time scale ($\tau_U$) 
associated with modified Urca processes, with the help of the expression 
for associated bulk viscosity $\zeta_U$ given by \cite{Sawyer}
\begin{equation}
 \zeta_U=1.46~\rho(r)^2\omega^{-2}
\left[\frac{k_B T}{1 MeV}\right]^6 {~\rm g/(cm\ s)}.
\label{bvUrca}
\end{equation}
The shear viscosity time scale is given by \cite{lom98}
\begin{equation}
\frac{1}{\tau_{\eta}} = \frac{(l-1)(2l+1)}{\int_0^R dr \rho(r) r^{2l+2}}
\int_0^R dr \eta r^{2l},\label{tauEta}
\end{equation}
where $\eta$ can be calculated from the prominent $nn$ scattering 
and is given by \cite{jai}
\begin{equation}
 \eta = 2\times 10^{18}\rho_{15}^{9/4}T_9^{-2} {~\rm g/(cm\ s)} .\label{Eta}
\end{equation}
Here $\rho_{15}=\rho/(10^{15}~{\rm g/cm^3})$ and $T_9=T/(10^9~{\rm K})$ 
are density and temperature respectively, casted in dimensionless forms.
Finally, the gravitational radiation time scale ($\tau_{GR}$) 
is given by \cite{lom98},
\begin{eqnarray}
 \frac{1}{\tau_{GR}}=&-&\frac{32 \pi G \Omega^{2l+2}}{c^{2l+3}}
\frac{(l-1)^{2l}}{\lbrack(2l+1)!!\rbrack^2} \left( \frac{l+2}{l+1}
 \right) ^{2l+2}
\nonumber \\ 
&\times& \int_0^R \rho(r) r^{2l+2} dr. \label{tauGR}
\end{eqnarray}

The evolution of the r-mode due to dissipative viscous effects and 
GR can be studied by defining the overall r-mode 
time scale $\tau_r$ \cite{lom98,LO},
\begin{equation}
{1\over\tau_r(\Omega,T)}={1\over\tau_{GR}(\Omega)}
+{1\over\tau_{B}(\Omega,T)}+{1\over\tau_{U}(\Omega,T)}
+{1\over\tau_{\eta}(\Omega,T)}
.\label{tau-r}
\end{equation}
It appears in the decay of the mode as $e^{-t/\tau_r}$ and 
when $\tau_r > 0$, the mode is stable. Now from Equation (\ref{tauGR}) 
we can see that $\tau_{GR} < 0$, which is indicative of the fact 
that GR allows the modes to grow and drives them to instability, 
while $\tau_B$, $\tau_U$ and $\tau_{\eta}$ are positive and thus 
they try to dampen the mode. We can define a critical angular velocity 
$\Omega_C$ as $1/\tau_r(\Omega_C,T)=0$; for a star at a
given temperature $T$. Now if the angular velocity of the star 
is greater than $\Omega_C$ then the star is unstable and will be 
subjected to GR emission while stars with angular velocities 
smaller than $\Omega_C$ will be stable.

\section{RESULTS AND DISCUSSIONS}

\begin{table}
\caption{Parameter set for the model.}
\vskip 0.1 in
\begin{center}
\begin{tabular}{cccccccc}
\hline
\hline
\multicolumn{1}{c}{$c_{\sigma N}$} &
\multicolumn{1}{c}{$c_{\omega N}$} &
\multicolumn{1}{c}{$c_{\rho N}$}   &
\multicolumn{1}{c}{$B$} &
\multicolumn{1}{c}{$C$} &
\multicolumn{1}{c}{$K$} &
\multicolumn{1}{c}{$m_N^{\star}/m_N$} \\
\multicolumn{1}{c}{($fm^2$)} &
\multicolumn{1}{c}{($fm^2$)} &
\multicolumn{1}{c}{($fm^2$)} &
\multicolumn{1}{c}{($fm^2$)} &
\multicolumn{1}{c}{($fm^4$)} &
\multicolumn{1}{c}{($MeV$)}  &
\multicolumn{1}{c}{}\\
\hline
6.79  &1.99  & 4.66  &-4.32  &0.165   &300  &0.85 \\
\hline
\end{tabular}
\end{center}
\end{table}

\begin{figure}
\includegraphics[width=7cm,height=7cm]{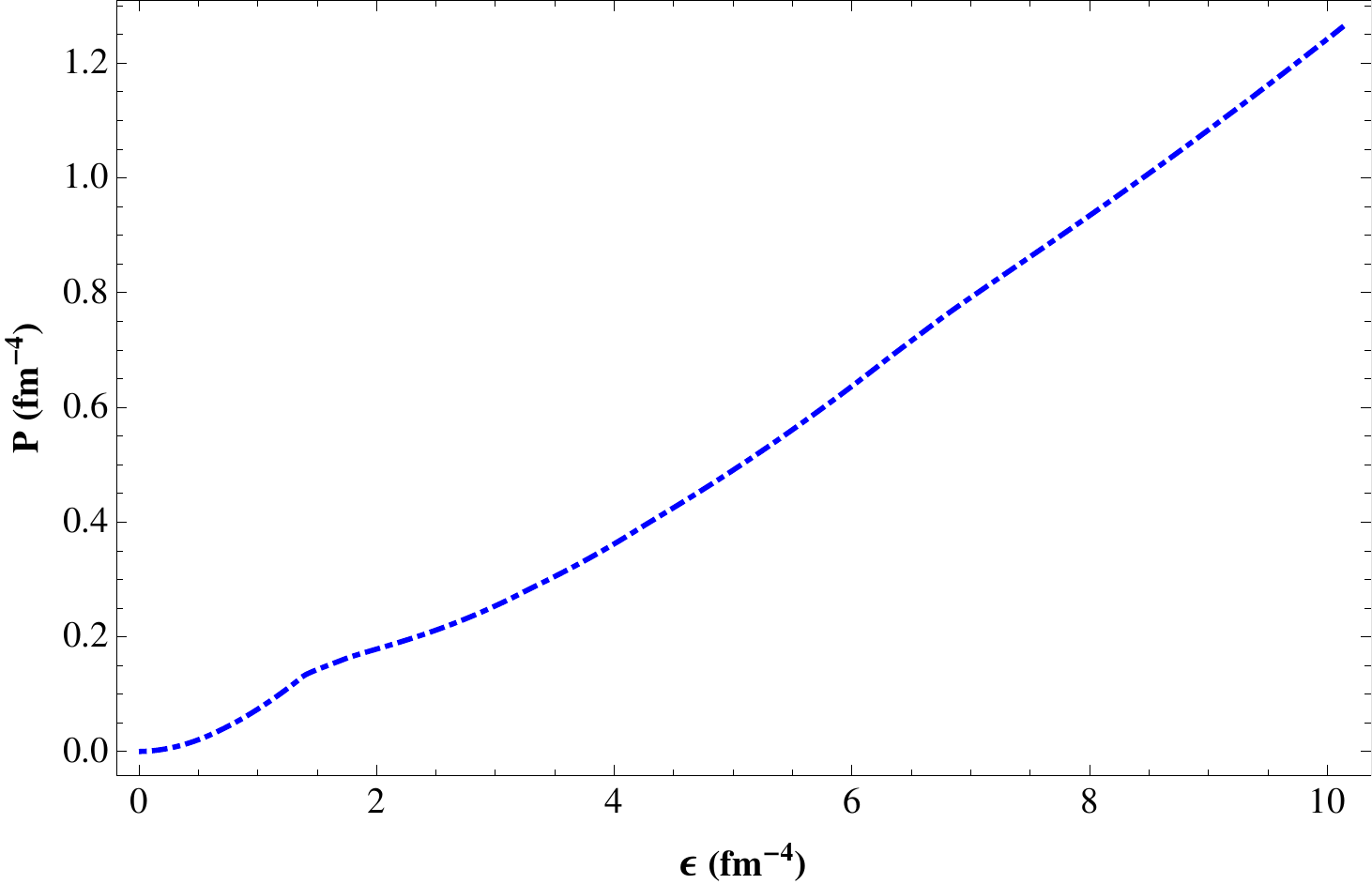}
\caption{Equation of State used to calculate the stellar configurations}\label{fig:1}
\end{figure}

\begin{figure}
\includegraphics[width=7.0cm,height=7.0cm]{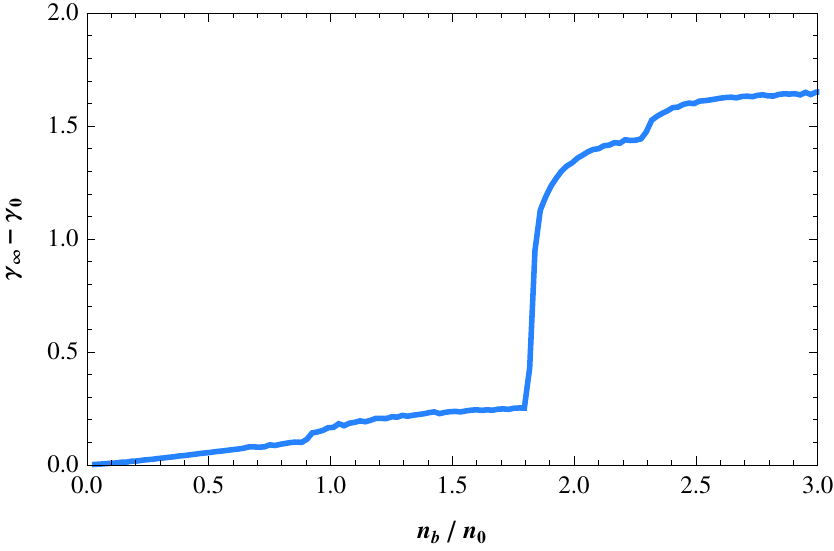}
\caption{Thermodynamic factor $\gamma_{\infty} - \gamma_0$ that appears in the expression for hyperon bulk viscosity, is plotted against the 
normalised baryon density.} \label{fig:2}
\end{figure}
Now with the EoS discussed in subsection \ref{eos} we set out to find 
the coefficient of bulk viscosity as given by equation (\ref{zeta}). The 
parameters in the effective chiral model that we have used are given in Table I.
The parameters were so chosen that they satisfy the constraints on the
equation of state from the flow data in heavy ion collisions \cite{tkj09}.
The resulting EOS is plotted in figure \ref{fig:1} .
To calculate the bulk-viscosity, we first need to calculate 
$\gamma_{\infty} - \gamma_0$, the difference between fast and slow 
adiabatic indices, from Eq.(\ref{gammas}). This expression can be
calculated with the help of EoS alone. In figure \ref{fig:2}, 
we plot $(\gamma_{\infty} - \gamma_0)$ as a function of the normalized baryon 
density ($n_{b}/n_0$), where $n_0$ = 0.153 $fm^{-3}$ is the nuclear matter
saturation density. The sudden rises in the graph can be attributed to 
the appearance of hyperons with increase of baryon density at the cost of 
neutron number density $n_n$.

\begin{figure}
\includegraphics[width=7cm,height=7cm]{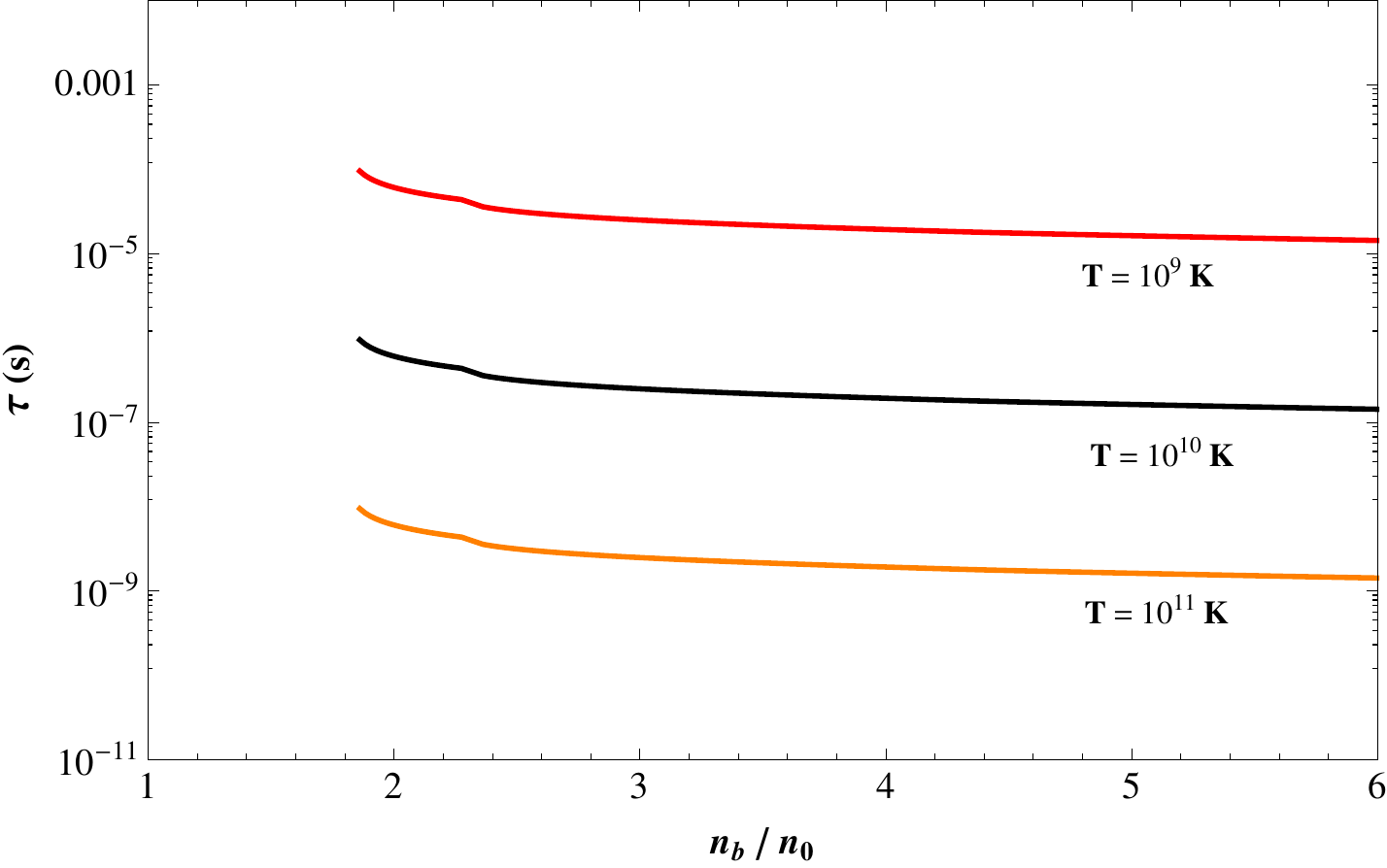}
\caption{Relaxation time $\tau$ (in seconds) for the non-leptonic 
processes causing hyperon bulk viscosity is plotted 
for various temperatures $T$}\label{fig:3}
\end{figure}
Since we have $\Sigma^-$ and $\Lambda$ hyperons formed in the system 
with lowest threshold densities, we consider the non leptonic reactions 
represented by the Eq. (\ref{r1}) and Eq.(\ref{r2}) and calculate
the relaxation time as given by the Eq. (\ref{tau}). We note that 
we have not considered the reaction Eq.(\ref{r3}) and further there 
are several reactions which are going to contribute to the net 
reaction rate \cite{dieperink,mohit}. The reason for this is that the rate
for the process given in Eq.(\ref{r3}) is estimated to be an order 
of magnitude higher than the process given by Eq.(\ref{r2}). This 
leads to subdominant contribution to the relaxation time and hence 
to the bulk viscosity \cite{pbjones}.
Thus what we are calculating is a lower limit of the net rate 
which will correspond to an upper limit on the bulk viscosity.
The matrix elements are calculated with the values of Fermi momenta
and effective masses of various baryonic species obtained from the EoS. 
Here we use axial-vector coupling values $g_{np}=-1.27,~g_{p\Lambda}=-0.72$ 
and $g_{n\Sigma}=0.34$ measured in $\beta$-decay of baryons at rest 
and Fermi coupling constant $G_F=1.166\times 10^{-11}$ MeV$^{-2}$ and 
sin$\theta_C=0.222$ (where $\theta_C$ is the Cabibbo weak mixing angle)
\cite{PDG}. Then, we calculate $\frac{\delta \mu}{n_B \delta x_{n}}$ from 
Eq. (\ref{delmu1}) for the densities where both the hyperons are present 
($n_B/n_0>2.36$), and, from Eq. (\ref{delmu2}) for lower densities where 
there is only $\Sigma$--hyperon present ($n_B/n_0=1.86-2.36$). 
Further, Eq.(\ref{alpha1})-Eq.(\ref{D}) are evaluated using the EoS
under consideration. We can thus calculate the relaxation time for 
relevant temperatures. We show the calculated behaviour of relaxation 
time (in seconds) with temperature in Fig. \ref{fig:3}. It is clear 
that the relaxation time increases considerably with the decrease 
of temperature. For a given temperature, the relaxation time is seen to
decrease when both the hyperons are present, as compared to the case 
of presence of a singe species of hyperons ($\Sigma$), since in this 
case $\tau$ value will be less according to Eq.(\ref{tau}). 
 \begin{figure}
\includegraphics[width=7cm,height=7cm]{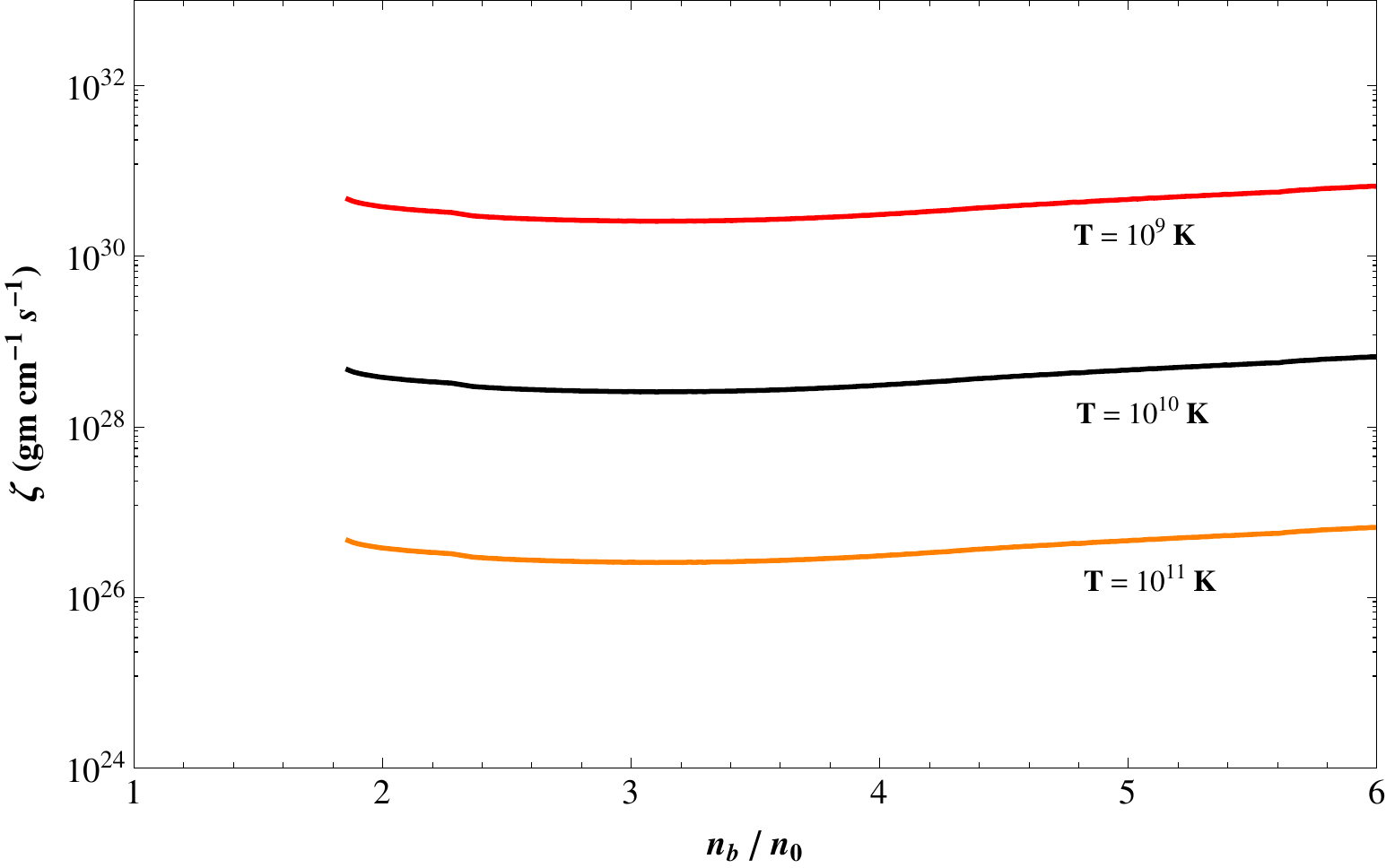}
\caption{Hyperon bulk viscosity $\zeta$ in units of g/(cm s) is plotted as a function of normalized baryon density for various 
temperatures}\label{fig:4}
\end{figure}
We then compute the coefficient of bulk viscosity responsible for
the mode damping in neutron stars from the expression (\ref{zeta}). 
The value of maximum frequency $\Omega_K$ is the Keplerian angular 
frequency of the rotating star and is set by the onset of mass shedding 
from the equator of the star. The bulk viscosity coefficient is 
calculated for the relevant temperatures 
and is plotted against the normalized baryon density in Fig. \ref{fig:4}. 
The behaviour of the hyperon bulk viscosity is similar to that of the 
corresponding relaxation time as is expected from Eq.(\ref{zeta}).
The high value of the bulk viscosity coefficient at the temperature 
$10^{9}$ K is indicative of the fact that hyperon 
bulk viscosity plays a major role in the suppression of the r-modes.
 We note that our bulk viscosity values are order of magnitude 
less than the values obtained by \cite{LO}. It could be  due to the 
fact that unlike their work we are not considering the effect of hyperon 
superfluidity in this calculation. It might also be noted that the 
non-superfluid hyperonic bulk viscosity calculated in \cite{ddprd06} uses 
an EoS based on a  model, where only $\Lambda$ hyperons are present 
at the relevant density.

\begin{figure}
\includegraphics[width=7cm,height=7cm]{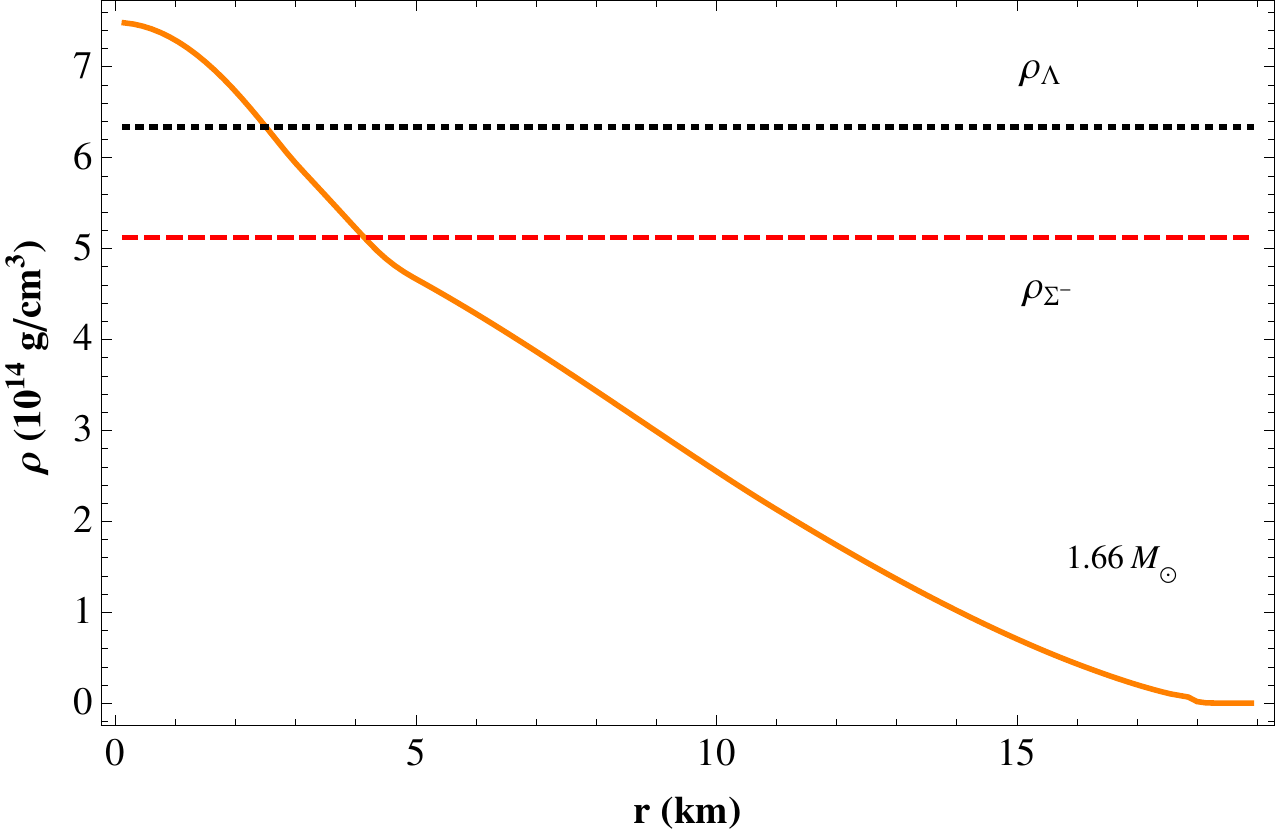}
\caption{Density of the star, $\rho$ (in units of $g/cm^3$), as a function 
of distance $r$ (in $km$) from the centre ($r$= 0) to the radius of the 
star ($r=~R$). Threshold densities corresponding to the formation of 
$\Sigma^-$ and $\Lambda^0$ hyperons are also plotted.} \label{fig:5}
\end{figure}
We next study the effect of hyperon bulk viscosity on the r-modes. 
Here we need to calculate the dissipation time scale 
due to hyperon bulk viscosity as well as due to other dissipative 
phenomena. If this time scale is greater than the gravitational 
radiation time scale, then the r mode is not stable and suppressed. 
In order to calculate 
the  dissipative timescales from Equations (\ref{taui})-(\ref{tauGR})
 we need to know 
the density profile $\rho(r)$, of the neutron star under consideration.
We need to know the Kepler frequency of the rotating star also. We use 
Tolman-Oppenheimer-Volkoff equations to construct the non-rotating 
stellar configurations. The maximum mass of the neutron star in this case 
is found to be $1.65~M_{\odot}$ with a radius of $16.7$ km. We use 
Hartle's slow rotation approximation to calculate the global properties of 
rotating neutron star\cite{hm-rot}. We get the maximum 
mass and radius ($R$) of the rotating star to be $1.66~M_{\odot}$ and 
$18.9$ km respectively. 
The Kepler frequency in this case is found to be $\Omega_K=3998 \rm~Hz.$ 
A typical density profile of the rotating star is shown in 
the Fig. \ref{fig:5} . This profile corresponds to a 
central density of $7.48 \times 10^{14}\rm~
gm/cm^3$. We have also indicated the densities corresponding to the 
appearance of the hyperons, i.e., threshold densities of both $\Sigma$ and 
$\Lambda$ hyperons in the graph. From centre of the star 
upto a density of $\rho=6.34 \times 10^{14}\rm~gm/cm^3$,
we have the presence of both the hyperons in the star 
(i.e., up to a distance $2.5$ km from the centre). 
The presence of $\Sigma$ alone is there upto 
$\rho=5.1 \times 10^{14}\rm~gm/cm^3$ (another $1.7$ km) 
making a hyperon core of radius $4.2$ km in the neutron star. 
Hyperon bulk viscosity time scale, and, hence its effects on r-mode 
is very sensitive to the hyperonic core's constituent structure 
and its radius.

\begin{figure}
\includegraphics[width=7cm,height=7cm]{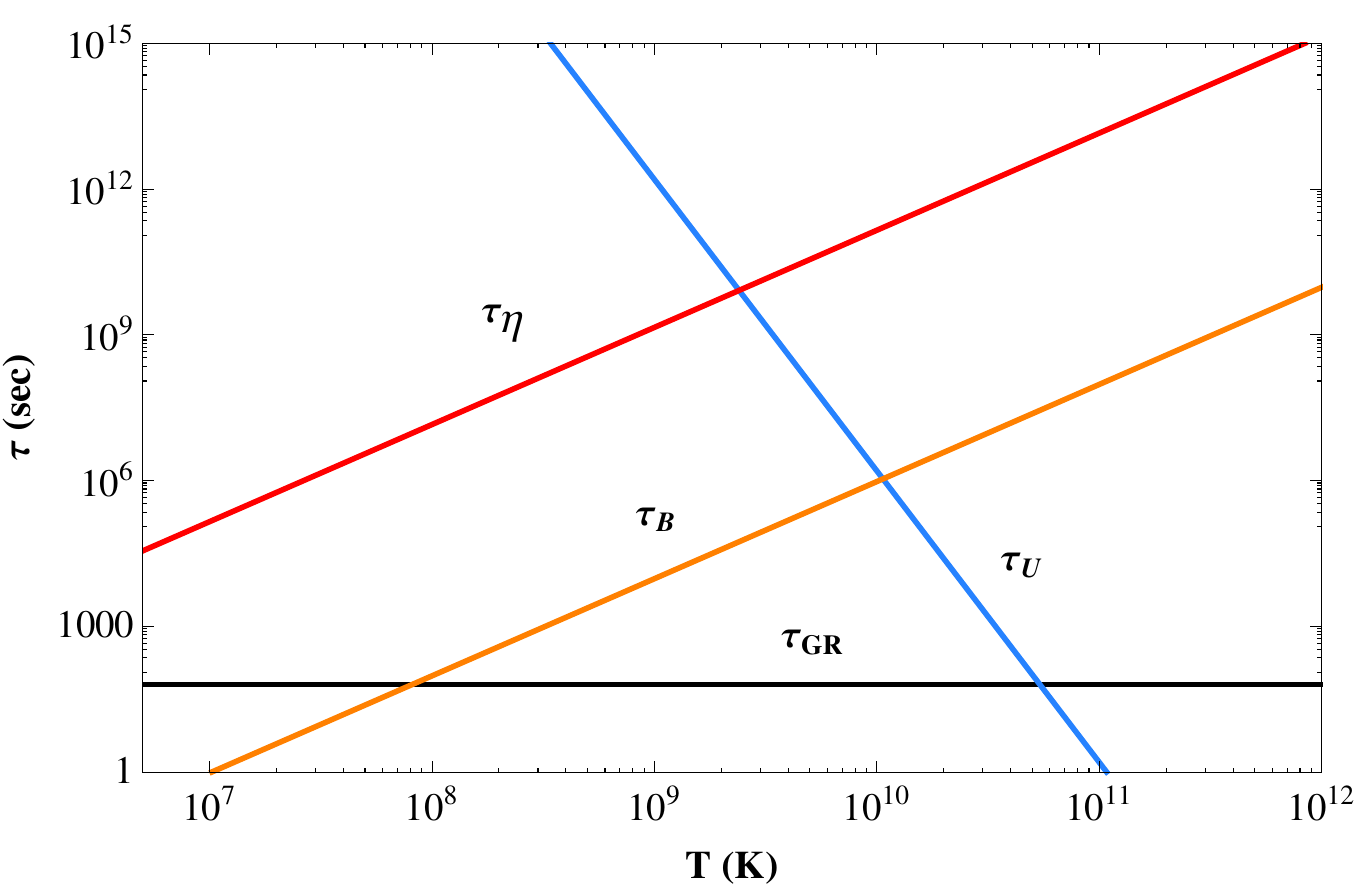}
\caption{The temperature dependence of damping time scales (in seconds) 
due to hyperonic bulk viscosity $\tau_{\zeta}$, 
\textit{modified} Urca bulk viscosity $\tau_{B}$ and shear viscosity 
$\tau_{\eta}$. $\tau_{GR}$ represents the temperature 
independent gravitational radiation time scale. 
(The star is considered to be rotating with the $Kepler$ frequency 
$\Omega_K$ here).}\label{fig:6}
\end{figure}

For the rotating neutron star with a mass of $1.66~M_{\odot}$ and 
$\Omega_K=3998 \rm~Hz$ as considered above, we next evaluate the 
various dissipative time scales associated with the r-mode damping. 
The dissipative time scale 
of hyperonic bulk viscosity, denoted by $\tau_B$ can be calculated from 
equations (\ref{taui})-(\ref{delv}) with the help of density profile of 
the star. Here hyperonic bulk viscosity ($\zeta$) as a function of radius 
is obtained from our previous calculations of $\zeta$ for the EoS together 
with the knowledge of stellar density profile i.e., $\zeta(\rho(r))$. 
The time scale associated with \textit{modified} Urca processes $\tau_U$, 
is calculated in the same manner as for $\tau_B$, by using the 
equation (\ref{bvUrca}) 
instead of hyperonic bulk viscosity. Next, we estimate the 
shear viscosity dissipative time scale $\tau_{\eta}$ using the 
equations (\ref{tauEta}) and (\ref{Eta}). Finally, the gravitational 
radiation time scale $\tau_{GR}$ 
associated with the r-mode can be calculated with the help of 
density profile using equation (\ref{tauGR}). Figure \ref{fig:6} 
shows the calculated time scales 
as functions of temperature, for the star rotating with $\Omega_K$.
From figure \ref{fig:6}, we observe that in the non-superfluid hyperonic 
matter, r-modes get substantially damped due to hyperonic bulk viscosity
 only at low temperatures $(T<10^8\rm~K)$, whereas the \textit{modified} 
Urca bulk viscosity suppresses the r-modes rapidly only at high 
temperatures $T>5 \times 10^{10}\rm~K$. The role of shear viscosity 
in suppressing the modes is not prominent in this temperature range. 
Consequently, the effect of r-mode instability will be prominent 
in the temperature window $(10^8 - 5 \times 10^{10})\rm~K$. 
The hyperon bulk viscosity
suppresses the instability for temperatures below $10^8$K while modified
Urca processes suppress the instability beyond $10^{10}K$.

\begin{figure}
\includegraphics[width=7cm,height=7cm]{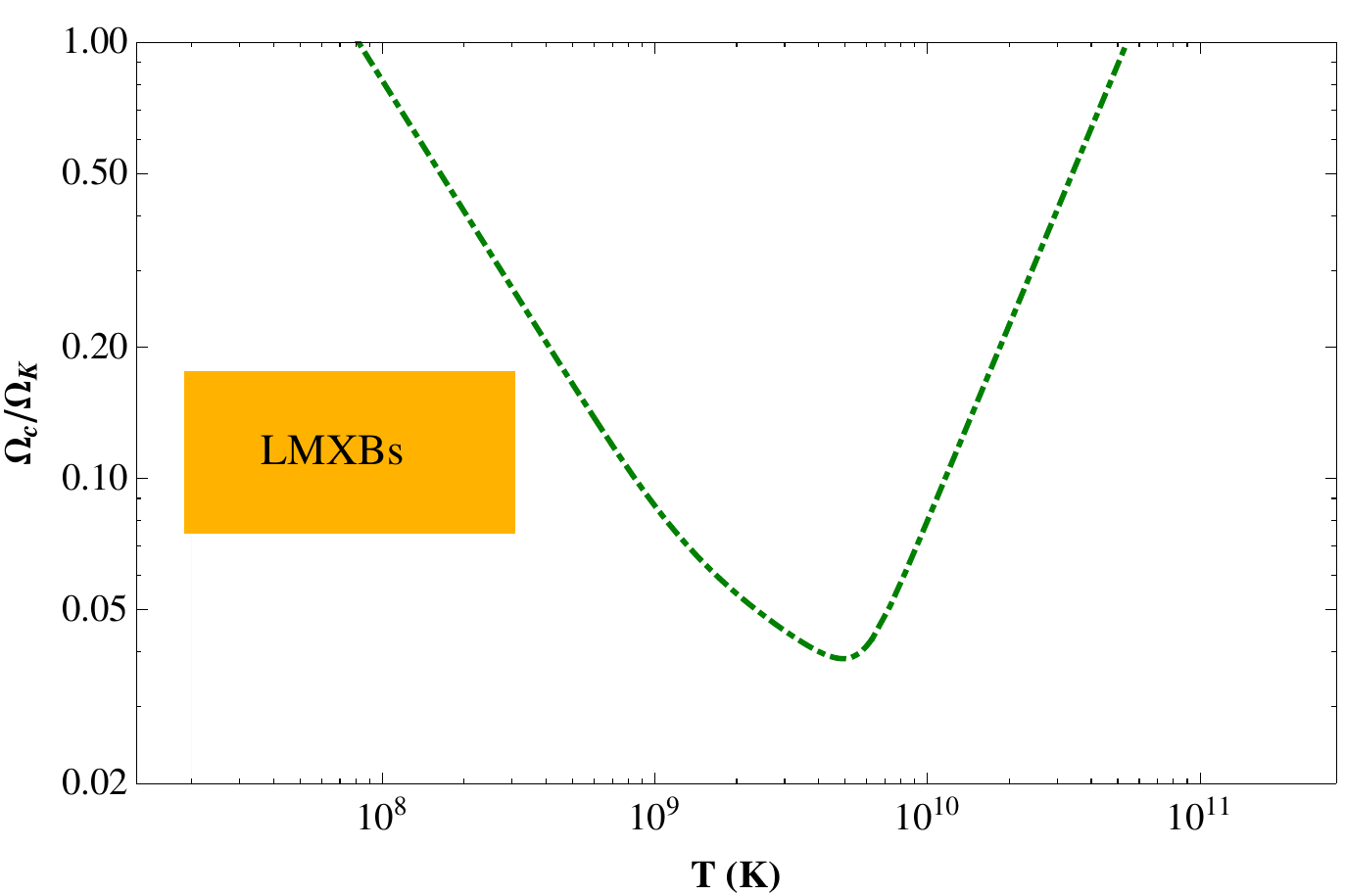}
\caption{Critical angular velocities (normalized to the $Kepler$ frequency 
$\Omega_K$ = 3998 $\rm Hz$) for a neutron star with mass of 
1.66 $M_{\odot}$ is shown as a function of hyperon core temperature. 
The shaded region represents the majority of the observed LMXBs.}\label{fig:7}
\end{figure}

Now we are in a position to calculate the critical angular velocity 
$\Omega_C$ of the neutron star. $\Omega_C$ is obtained by solving 
equation (\ref{tau-r}); $\frac{1}{\tau_r(\Omega_C,T)}=0$, for a 
particular value of $T$. At this frequency, the energy fed 
into the r-mode per unit time by gravitational radiation is equal 
to the energy dissipated per unit time. A star rotating above this 
critical frequency will be subjected to r-mode instability.
We have shown the $\Omega_C$ (scaled to $\Omega_K$) 
in Fig. \ref{fig:7} for the temperature regime of interest.
Since $\Omega_K$ determines the maximum allowed rotation rate for the 
star, stable rotation at any temperature will have $\Omega_C/\Omega_K=1$ 
as an upper bound. In this figure, the region above $\Omega_C$ curve 
is unstable and a star rotating in this region will be rapidly spun 
down to an angular frequency below $\Omega_C$. As expected the 
instability window exists in the temperature regime 
$(10^8 - 5 \times 10^{10})\rm~K$, where gravitational radiation 
is dominant and not suppressed substantial, which shows that the
neutron star with hyperonic core is unstable in this region. 
Low temperature regime hyperonic bulk viscosity damps the r-mode 
effectively whereas nucleon dominated \textit{modified} Urca bulk 
viscosity is the cause of mode damping at high temperatures. 
The minima of $\Omega_C$ curve occurs at $T \approx 5 \times 10^{10} 
\rm~K$ with $\Omega_C\approx.04~\Omega_K$, which is indicative of 
the fact that the r-mode instability is rather strong in this 
hyperon core scenario. The shaded box in the Fig. \ref{fig:7} 
is where most of the observed Low Mass 
X-ray Binaries (LMXBs) are found. They have a core temperature in 
the range of $(2 \times10^7 - 3 \times 10^8)\rm~K$ with rotation 
rate between $300$ to $700 \rm~Hz$ \cite{jai,LMXB}. In our case,
LMXBs are placed in the stable region, unlike conventional neutron 
stars with $npe$ matter \cite{basil}.

\section{SUMMARY}
Rotating equilibrium configurations of self gravitating fluids are 
subjected to various possibilities of instabilities at large rotation 
periods. In the present work, we have investigated the r-mode instability 
which is known to limit the angular velocities of rapidly rotating stars. 
The r-mode and the 
related instabilities are damped by various viscosities of the matter
in the interior of the neutron star. Thus the microscopic models 
describing the matter in the interior of the star get constrained by the 
observations of the rapidly rotating pulsars.

In the present work, we have confined our attention to the case of 
neutron star with a hyperonic core. For the description of the
matter in the core of the neutron star, we have used an effective
chiral hadronic model generalised to include the lowest lying octet
of baryons. The parameters of the model are chosen that are consistent 
with the flow data in heavy ion collisions, nuclear matter properties 
as well as observation of high masss neutron stars. In the present work,
we have computed the coefficient of bulk viscosity due to the the hyperonic 
matter in the core of a neutron star and the resulting effects on the 
r-mode instability. It turns out that hyperon bulk viscosity within 
the model is effective in damping the instability for temperatures 
below $10^8$K . Beyond a temperature of about $10^{10}$K, the bulk 
viscosity due to modified Urca processes become effective
in damping the r-mode instability. Shear viscosity of hadronic matter
becomes effective in damping only at low temperatures. 
Within the model it turns out that the the  bulk viscosity in normal 
hyperonic matter does play an important role in spinning down fast 
rotating neutron stars. However, superfluid hyperonic matter or 
quark matter in the core can change this conclusion.

We have not considered in the present work the phase transition to quark
matter which could be most likely in a color superconducting
phase. The role of quark matter in the context of r-mode characteristics
has been dealt with in Ref.\cite{jai} which shows that the r-mode 
instability gets suppressed by both normal quark matter as well as
gapped quark matter in the color flavor locked phase. 
Further, the neutron stars are also endowed with strong magnetic fields
the effects of which on the r-modes have not been included in the 
present work. Ultra strong magnetic field seem to increase the 
instability window for quark matter \cite{rischkehuang}. It will 
thus be interesting to examine the scenario with a phase transition 
to quark matter and the effect of magnetic fields in the context 
of r-modes for a hybrid star with a crust of hadronic matter. 
Work in this direction is under progress and will be reported elsewhere
\cite{future}.
 
\acknowledgements The authors would like to thank Jitesh R Bhatt,
Debades Bandyopadhyay and Debarati Chatterjee for many discussions. 
The authors would also like to thank Amruta Mishra 
for a careful reading of the manuscript.
 
\def \ltg{R.P. Feynman, Nucl. Phys. B 188, 479 (1981); 
K.G. Wilson, Phys. Rev. \zbf  D10, 2445 (1974); J.B. Kogut,
Rev. Mod. Phys. \zbf  51, 659 (1979); ibid  \zbf 55, 775 (1983);
M. Creutz, Phys. Rev. Lett. 45, 313 (1980); ibid Phys. Rev. D21, 2308
(1980); T. Celik, J. Engels and H. Satz, Phys. Lett. B129, 323 (1983)}

\def\loff{A.I. Larkin and Yu.N. Ovchinnikov, Sov. Phys. JETP{\bf 20} (1965);
P. Fulde and R.A. Ferrel, Phys Rev. {\bf A135}, 550, 1964.}
\def\takada{S. Takada and T. Izuyama, Prog. theor. Phys. {\bf 41}, 635 (1969).}
\def\ren{I. Giannakis, H. Ren,{\PLB{611}{137}{2005}};
{\em ibid}{\NPB{723}{255}{2005}}.}
\def\berges {J. Berges, K. Rajagopal, {\NPB{538}{215}{1999}}.}
\def \svz {M.A. Shifman, A.I. Vainshtein and V.I. Zakharov,
Nucl. Phys. B147, 385, 448 and 519 (1979);
R.A. Bertlmann, Acta Physica Austriaca 53, 305 (1981)}

\def \spmbst {S.P. Misra, Phys. Rev. D35, 2607 (1987)}

\def \hmgrnv { H. Mishra, S.P. Misra and A. Mishra,
Int. J. Mod. Phys. A3, 2331 (1988)}

\def \snss {A. Mishra, H. Mishra, S.P. Misra
and S.N. Nayak, Phys. Lett 251B, 541 (1990)}

\def \amqcd { A. Mishra, H. Mishra, S.P. Misra and S.N. Nayak,
Pramana (J. of Phys.) 37, 59 (1991). }
\def\qcdtb{A. Mishra, H. Mishra, S.P. Misra 
and S.N. Nayak, Z.  Phys. C 57, 233 (1993); A. Mishra, H. Mishra
and S.P. Misra, Z. Phys. C 58, 405 (1993)}

\def \spmtlk {S.P. Misra, Talk on {\it `Phase transitions in quantum field
theory'} in the Symposium on Statistical Mechanics and Quantum field theory, 
Calcutta, January, 1992, hep-ph/9212287}

\def \hmnj {H. Mishra and S.P. Misra, 
{\PRD{48}{5376}{1993}.}}

\def \hmqcd {A. Mishra, H. Mishra, V. Sheel, S.P. Misra and P.K. Panda,
hep-ph/9404255 (1994)}

\def \amcrl {A. Mishra, H. Mishra and S.P. Misra, Z. Phys. C 57, 241 (1993)}

\def \higgs { S.P. Misra, in {\it Phenomenology in Standard Model and Beyond}, 
Proceedings of the Workshop on High Energy Physics Phenomenology, Bombay,
edited by D.P. Roy and P. Roy (World Scientific, Singapore, 1989), p.346;
A. Mishra, H. Mishra, S.P. Misra and S.N. Nayak, Phys. Rev. D44, 110 (1991)}

\def \nmtr {A. Mishra, 
H. Mishra and S.P. Misra, Int. J. Mod. Phys. A5, 3391 (1990); H. Mishra,
 S.P. Misra, P.K. Panda and B.K. Parida, Int. J. Mod. Phys. E 1, 405, (1992);
 {\it ibid}, E 2, 547 (1993); A. Mishra, P.K. Panda, S. Schrum, J. Reinhardt
and W. Greiner, to appear in Phys. Rev. C}

\def \dtrn {P.K. Panda, R. Sahu and S.P. Misra, 
Phys. Rev C45, 2079 (1992)}

\def \qcd {G. K. Savvidy, Phys. Lett. 71B, 133 (1977);
S. G. Matinyan and G. K. Savvidy, Nucl. Phys. B134, 539 (1978); N. K. Nielsen
and P. Olesen, Nucl.  Phys. B144, 376 (1978); T. H. Hansson, K. Johnson,
C. Peterson Phys. Rev. D26, 2069 (1982)}

\def \cornwal {J.M. Cornwall, Phys. Rev. D26, 1453 (1982)}
\def\aichlin {F. Gastineau, R. Nebauer and J. Aichelin,
{\PRC{65}{045204}{2002}}.}

\def \mndglv {J. E. Mandula and M. Ogilvie, Phys. Lett. 185B, 127 (1987)}

\def \schwinger {J. Schwinger, Phys. Rev. 125, 1043 (1962); ibid,
127, 324 (1962)}

\def \schutte {D. Schutte, Phys. Rev. D31, 810 (1985)}

\def \amspm {A. Mishra and S.P. Misra, 
{\ZPC{58}{325}{1993}}.}

\def \gft{ For gauge fields in general, see e.g. E.S. Abers and 
B.W. Lee, Phys. Rep. 9C, 1 (1973)}

\def \gribov {V.N. Gribov, Nucl. Phys. B139, 1 (1978)}

\def \spm78 {S.P. Misra, Phys. Rev. D18, 1661 (1978); {\it ibid}
D18, 1673 (1978)} 

\def \lopr {A. Le Youanc, L.  Oliver, S. Ono, O. Pene and J.C. Raynal, 
Phys. Rev. Lett. 54, 506 (1985)}

\def \spphi {S.P. Misra and S. Panda, Pramana (J. Phys.) 27, 523 (1986);
S.P. Misra, {\it Proceedings of the Second Asia-Pacific Physics Conference},
edited by S. Chandrasekhar (World Scientific, 1987) p. 369}

\def\spmdif {S.P. Misra and L. Maharana, Phys. Rev. D18, 4103 (1978); 
    S.P. Misra, A.R. Panda and B.K. Parida, Phys. Rev. Lett. 45, 322 (1980);
    S.P. Misra, A.R. Panda and B.K. Parida, Phys. Rev. D22, 1574 (1980)}

\def \spmvdm {S.P. Misra and L. Maharana, Phys. Rev. D18, 4018 (1978);
     S.P. Misra, L. Maharana and A.R. Panda, Phys. Rev. D22, 2744 (1980);
     L. Maharana,  S.P. Misra and A.R. Panda, Phys. Rev. D26, 1175 (1982)}

\def\spmthr {K. Biswal and S.P. Misra, Phys. Rev. D26, 3020 (1982);
               S.P. Misra, Phys. Rev. D28, 1169 (1983)}

\def \spmstr { S.P. Misra, Phys. Rev. D21, 1231 (1980)} 

\def \spmjet {S.P. Misra, A.R. Panda and B.K. Parida, Phys. Rev Lett. 
45, 322 (1980); S.P. Misra and A.R. Panda, Phys. Rev. D21, 3094 (1980);
  S.P. Misra, A.R. Panda and B.K. Parida, Phys. Rev. D23, 742 (1981);
  {\it ibid} D25, 2925 (1982)}

\def \arpftm {L. Maharana, A. Nath and A.R. Panda, Mod. Phys. Lett. 7, 
2275 (1992)}

\def \van {R. Van Royen and V.F. Weisskopf, Nuov. Cim. 51A, 617 (1965)}

\def \rchpi {S.R. Amendolia {\it et al}, Nucl. Phys. B277, 168 (1986)}

\def \chrl{ Y. Nambu, {\PRL{4}{380}{1960}};
A. Amer, A. Le Yaouanc, L. Oliver, O. Pene and
J.C. Raynal,{\PRL{50}{87}{1983a}};{\em ibid}
{\PRD{28}{1530}{1983}};
M.G. Mitchard, A.C. Davis and A.J.
MAacfarlane, {\NPB{325}{470}{1989}};
B. Haeri and M.B. Haeri,{\PRD{43}{3732}{1991}};
V. Bernard,{\PRD{34}{1604}{1986}};
 S. Schramm and
W. Greiner, Int. J. Mod. Phys. \zbf E1, 73 (1992), 
J.R. Finger and J.E. Mandula, Nucl. Phys. \zbf B199, 168 (1982),
S.L. Adler and A.C. Davis, Nucl. Phys.\zbf  B244, 469 (1984),
S.P. Klevensky, Rev. Mod. Phys.\zbf  64, 649 (1992).}

\def \spmijp { S.P. Misra, Ind. J. Phys. 61B, 287 (1987)}

\def \feynman {R.P. Feynman and A.R. Hibbs, {\it Quantum mechanics and
path integrals}, McGraw Hill, New York (1965)}

\def \glstn{ J. Goldstone, Nuov. Cim. \zbf 19, 154 (1961);
J. Goldstone, A. Salam and S. Weinberg, Phys. Rev. \zbf  127,
965 (1962)}

\def \anderson {P.W. Anderson, Phys. Rev. \zbf {110}, 827 (1958)}

\def \nambu{ Y. Nambu, Phys. Rev. Lett. \zbf 4, 380 (1960)}

\def\donogh {J.F. Donoghue, E. Golowich and B.R. Holstein, {\it Dynamics
of the Standard Model}, Cambridge University Press (1992)}

\def\satz {T. Matsui and H. Satz, Phys. Lett. B178, 416 (1986)}

\def\cps {C. P. Singh, Phys. Rep. 236, 149 (1993)}

\def\prliop {A. Mishra, H. Mishra, S.P. Misra, P.K. Panda and Varun
Sheel, Int. J. of Mod. Phys. E 5, 93 (1996)}

\def\hmcor {V. Sheel, H. Mishra and J.C. Parikh, Phys. Lett. B382, 173
(1996); {\it biid}, to appear in Int. J. of Mod. Phys. E}
\def\cort { V. Sheel, H. Mishra and J.C. Parikh, Phys. ReV D59,034501 (1999);
{\it ibid}Prog. Theor. Phys. Suppl.,129,137, (1997).}

\def\surcor {E.V. Shuryak, Rev. Mod. Phys. 65, 1 (1993)} 

\def\stevenson {A.C. Mattingly and P.M. Stevenson, Phys. Rev. Lett. 69,
1320 (1992); Phys. Rev. D 49, 437 (1994)}

\def\mac {M. G. Mitchard, A. C. Davis and A. J. Macfarlane,
 Nucl. Phys. B 325, 470 (1989)} 
\def\tfd
 {H.~Umezawa, H.~Matsumoto and M.~Tachiki {\it Thermofield dynamics
and condensed states} (North Holland, Amsterdam, 1982) ;
P.A.~Henning, Phys.~Rep.253, 235 (1995).}
\def\amph4{Amruta Mishra and Hiranmaya Mishra,
{\JPG{23}{143}{1997}}.}

\def \neglecor{M.-C. Chu, J. M. Grandy, S. Huang and 
J. W. Negele, Phys. Rev. D48, 3340 (1993);
ibid, Phys. Rev. D49, 6039 (1994)}

\def\revdata {Particle Data Group, Phys. Rev. D 50, 1173 (1994)}

\def\sinp {S.P. Misra, Indian J. Phys., {\bf 70A}, 355 (1996)}
\def\hmparikh{H. Mishra and J.C. Parikh, {\NPA{679}{597}{2001}.}}
\def\krisch {M. Alford and K. Rajagopal, JHEP 0206,031,(2002)}
\def\reddy {
A.W. Steiner, S. Reddy and M. Prakash, {\PRD{66}{094007}{2002}.}}
\def\hmam {Amruta Mishra and Hiranmaya Mishra,
{\PRD{69}{014014}{2004}.}}
\def\hmampp {Amruta Mishra and Hiranmaya Mishra,
in preparation.}
\def\bryman {D.A. Bryman, P. Deppomier and C. Le Roy, Phys. Rep. 88,
151 (1982)}
\def\thooft {G. 't Hooft, Phys. Rev. D 14, 3432 (1976); D 18, 2199 (1978);
S. Klimt, M. Lutz, U. Vogl and W. Weise, Nucl. Phys. A 516, 429 (1990)}
\def\alkz { R. Alkofer, P. A. Amundsen and K. Langfeld, Z. Phys. C 42,
199(1989), A.C. Davis and A.M. Matheson, Nucl. Phys. B246, 203 (1984).}
\def\sarah {T.M. Schwartz, S.P. Klevansky, G. Papp,
{\PRC{60}{055205}{1999}}.}
\def\wil{M. Alford, K.Rajagopal, F. Wilczek, {\PLB{422}{247}{1998}};
{\it{ibid}}{\NPB{537}{443}{1999}}.}
\def\sursc{R.Rapp, T.Schaefer, E. Shuryak and M. Velkovsky,
{\PRL{81}{53}{1998}};{\it ibid}{\AP{280}{35}{2000}}.}
\def\pisarski{
D. Bailin and A. Love, {\PR{107}{325}{1984}},
D. Son, {\PRD{59}{094019}{1999}}; 
T. Schaefer and F. Wilczek, {\PRD{60}{114033}{1999}};
D. Rischke and R. Pisarski, {\PRD{61}{051501}{2000}}, 
D. K. Hong, V. A. Miransky, 
I. A. Shovkovy, L.C. Wiejewardhana, {\PRD{61}{056001}{2000}}
.}
\def\leblac {M. Le Bellac, {\it Thermal Field Theory}(Cambridge, Cambridge University
Press, 1996).}
\def\bcs{A.L. Fetter and J.D. Walecka, {\it Quantum Theory of Many
particle Systems} (McGraw-Hill, New York, 1971).}
\def\alexander{Aleksander Kocic, Phys. Rev. D33, 1785,(1986).}
\def\bubmix{F. Neumann, M. Buballa and M. Oertel,
{\NPA{714}{481}{2003}.}}
\def\kunihiro{M. Kitazawa, T. Koide, T. Kunihiro, Y. Nemeto,
{\PTP{108}{929}{2002}.}}
\def\igor{Igor Shovkovy, Mei Huang, {\PLB{564}{205}{2003}}.}
\def\prasanth{P. Jaikumar and M. Prakash,{\PLB{516}{345}{2001}}.}
\def\igorr{Mei Huang, Igor Shovkovy, {\NPA{729}{835}{2003}}.}
\def\abrikosov{A.A. Abrikosov, L.P. Gorkov, Zh. Eskp. Teor.39, 1781,
1960}
\def\krischprl{M.G. Alford, J. Berges and K. Rajagopal,
 {\PRL{84}{598}{2000}.}}
\def\hatmampp{A. Mishra and H.Mishra, in preparation}
\def\blaschke{D. Blaschke, M.K. Volkov and V.L. Yudichev,
{\EPJA{17}{103}{2003}}.}
\def\mei{M. Huang, P. Zhuang, W. Chao,
{\PRD{65}{076012}{2002}}}
\def\bubnp{
M. Buballa, M. Oertel, {\NPA{703}{770}{2002}}.}
\def\sarma{G. Sarma, J. Phys. Chem. Solids 24,1029 (1963).}
\def\ebert {D. Ebert, H. Reinhardt and M.K. Volkov,
Prog. Part. Nucl. Phys.{\bf 33},1, 1994.}
\def\rehberg{ P. Rehberg, S.P. Klevansky and J. Huefner,
{\PRC{53}{410}{1996}.}}
\def\lutz{M. Lutz, S. Klimt, W. Weise,{\NPA{542}{521}{1992}.}}
\def\rapid{B. Deb, A.Mishra, H. Mishra and P. Panigrahi,
Phys. Rev. A {\bf 70},011604(R), 2004.}
\def\kriscfl{M. Alford, C. Kouvaris, K. Rajagopal, Phys. Rev. Lett.
{\bf 92} 222001 (2004), arXiv:hep-ph/0406137.}
\def\shovris{S.B. Ruester, I.A. Shovkovy and D.H. Rischke,
arXiv:hep-ph/0405170.}
\def\krisaug{K. Fukushima, C. Kouvaris and K. Rajagopal, arxiv:hep-ph/0408322}.
\def\wilczek{W.V. Liu and F. Wilczek,{\PRL{90}{047002}{2003}},E. Gubankova,
W.V. Liu and F. Wilczek, {\PRL{91}{032001}{2003}.}}
\def\colscrev{For reviews see
M.G. Alford, A. Schmitt, K. Rajagopal and T. Schaefer, arXiv:0709.4635
 K. Rajagopal and F. Wilczek,
arXiv:hep-ph/0011333; D.K. Hong, Acta Phys. Polon. B32,1253 (2001);
M.G. Alford, Ann. Rev. Nucl. Part. Sci 51, 131 (2001); G. Nardulli,
Riv. Nuovo Cim. 25N3, 1 (2002); S. Reddy, Acta Phys Polon.B33, 4101(2002);
T. Schaefer arXiv:hep-ph/0304281; D.H. Rischke, Prog. Part. Nucl. Phys. 52,
197 (2004); H.C. Ren, arXiv:hep-ph/0404074; M. Huang, arXiv: hep-ph/0409167;
I. Shovkovy, arXiv:nucl-th/0410191.}
\def\kunihiroo{ M. Kitazawa, T. Koide, T. Kunihiro and Y. Nemoto,
{\PRD{65}{091504}{2002}}, D.N. Voskresensky, arXiv:nucl-th/0306077.}
\def\rupak{S.Reddy and G. Rupak, arXiv:nucl-th/0405054}
\def\ida{K. Iida and G. Baym,{\PRD{63}{074018}{2001}},
Erratum-ibid{\PRD{66}{059903}{2002}}; K. Iida, T. Matsuura, M. Tachhibana 
and T. Hatsuda, {\PRL{93}{132001}{2004}}; ibid,{arXiv:hep-ph/0411356}}
\def\chromo{Mei Huang and Igor Shovkovy,{\PRD{70}{051501}{2004}};
 {\em ibid}, {\PRD{70}{094030}{2004}}}
\def\steiner{A.W. Steiner, {\PRD{72}{054024}{2005}.}}
\def\andreaskris{K. Rjagopal and A. Schimitt{\PRD{73}{045003}{2006}.}}
\def\andreas{J.Deng, A. Schmitt and Q. Wang {\PRD{76}{034013}{2007}.}}
\def\deng{J.Deng, J. Wang and Q. Wang arXiv:0803.4360.}
\def\krisandreas{K. Rajagopal and A. Schmitt, {\PRD{73}{045003}{2006}.}}
\def\abuki{H. Abuki,{\NPA{791}{117}{2007}}.}
\def\abukib{Y. Nishida and H. Abuki,{\PRD{72}{096004}{2005}.}}
\def\brauner{Tomas Brauner, arXiv:0803.2422[hep-ph].}
\def\nardulli{M. Mannarelli, G. Nardulli and M. Ruggieri, {\PRA{74}{033606}{2006}}.}
\def\rischke{M. Kitazawa, D. Rischke and I.A. Shovkovy, arXiv:0709.2235 hep-ph.}
\def\amhm5{A. Mishra and H. Mishra, {\PRD{71}{074023}{2005}.}}
\def\leupold{K. Schertler, S. Leupold and J. Schaffner-Bielich,
{\PRC{60}{025801}(1999).}}
\def\bubrep{Michael Buballa, Phys. Rep.{\bf 407},205, 2005.}
\def\hatkun{T. Hatsuda and T. Kunihiro, Phys. Rep.{\bf 247},221, 1994.}
\def\hatsuda{H. Abuki, T. Hatsuda, K. Itakura, {\PRD{65}{074014}{2002}}.}
\def\lkw{ M. Lutz, S. Klimt and W. Weise, Nucl Phys. {\bf A542}, 521, 1992.}
\def\ruester{S.B. Ruester, V.Werth, M. Buballa, I. Shovkovy, D.H. Rischke,
arXiv:nucl-th/0602018;
 S.B. Ruester, I. Shovkovy, D.H. Rischke, {\NPA{743}{127}{2004}.}}
\def\hmparikh{H. Mishra and J.C. Parikh, {\NPA{679}{597}{2001}.}}
\def\amhma{Amruta Mishra and Hiranmaya Mishra,
{\PRD{69}{014014}{2004}.}}
\def\amhmb{A. Mishra and H. Mishra, {\PRD{71}{074023}{2005}.}}
\def\amhmc{A. Mishra and H. Mishra, {\PRD{74}{054024}{2006}.}}
\def\caldas{H. Caldas, arXiv:cond-mat/0605005}
\def\amhmloff{A. Mishra and H. Mishra, arXiv:cond-mat/0611058.}
\def\randeria{C.A.R. Sa de Melo, M. Randeria and J.R. Engelbrecht,
{\PRL{71}{3202}{1993}}.}
\def\carlsonreddy{S.Y. Chang, J. Carlson, V.R. Pandharipande and K.E. Schmidt,
{\PRA{70}{043602}{2004}}; J. Carlson and S. Reddy, {\PRL{95}{060401}{2005}}.}
\def\nishidason{Y. Nishida and D.T. Son, {\PRL{97}{050403}{2006}},
G. Rupak, T. Schafer and A. Kryjevski, {\PRA{75}{023606}{2007}}.}
\def\zhuang{G.F. Sun, L.He, P. Zhuang, {\PRD{75}{096004}{2007}}, 
L. He, P. Zhuang {\PRD{76}{056003}{2007}}.}
\def\zhuangb{L.He, P. Zhuang, {\PRD{75}{096003}{2007}}.}
\def\chennakano{J.W. Chen and E. Nakano, {\PRA{75}{043620}{2007}}.}
\def\rishi{K.Rajagopal and R. Sharma, {\PRD{74}{094019}{2006}}}
\def\cameliapi{G. Amelino-Camelia and S.Y. Pi,{\PRD{47}{2356}{1993}.}}
\def\rislena{D. Rischke and J. Lenaghan,{\JPG{26}{431}{2000}.}}
\def\risph{I. Giannakis, D.Hou and H.Ren and 
D. Rischke,{\PRL{93}{232301}{2004}}.}
\def\pirner{A.H. Rezaean and H.J. Pirner, {\NPA{779}{197}{2006}}}
\def\hypstar{N.K. Glendenning and S.A. Moskowski, {\PRL{67}{2414}{1991}},
J. Scahaffner and I.N. Mishustin, {\PRC{53}{1416}{1996}},M. Prakash,
I Bombaci, M. Prakash, P.J. Ellis, J. M. Lattimer and R. Knorren,
{\PR{280}{1}{1997}}}
\def\weber{F. Weber, Prog. Part. Nucl. Phys.{\bf 54}, 193,2005.}
\def\qmatter{N. Itoh, prog. Theor. Phys. {\bf 44}, 291,(1975), 
J.C. Collins and M.J. Perry, {\PRL{34}{1353}{1975}}.}
\def\witten{E. Witten,{\PRD{30}{272}{1984}}.}
\def\chargen{
 S.B. Ruester, I. Shovkovy, D.H. Rischke, {\NPA{743}{127}{2004}.};
K. Rajagopal and A. Schmitt, {\PRD{73}{045003}{2006}};
M. Buballa, M. Oertel, {\NPA{703}{770}{2002}};
A.W. Steiner, S. Reddy and M. Prakash, {\PRD{66}{094007}{2002}.}}
\def\hmamchargen{
Amruta Mishra and Hiranmaya Mishra,
\PRD{69}{014014}{2004};ibid,
{\PRD{71}{074023}{2005}};ibid,
{\PRD{74}{054024}{2006}.}}
\def\rmoderev{N. Andersson and K.D. Kokotas, Int. J. mod. Phys. {\bf D10},
381, (2001); L. Lindblom, arXiv:astro-ph/0101136.}
\def\cfs{S. Chandrasekhar,{\PRL{24}{611}{1970}}; 
J.L. Friedman and B.F. Schutz, Astrophys. J. {\bf 222},281(1978).}
\def\drago{A. Drago, A. Lavagno and G. Pagliara {\PRD{71}{103004}{2005}};
A. Drago, G. Pagliara and I. Parenti{\PRD{75}{123506}{2007}}}
\def\jones{P.B. Jones,{\PRL{86}{1384}{2001}};ibid {\PRD{64}{084003}{2001}}.}
\def\pbjones{P.B. Jones,{\PRD{64}{084003}{2001}}.}
\def\deba{D. Chatterjee and D. Bandyopadhyay,{\PRD{74}{023003}{2006}};
ibid,{\PRD{75}{123006}{2007}}}.
\def\dieperink{E.N.E. van Dalen and A.E.L. Dieperink,{\PRC{69}{025802}{2004}}.}
\def\linmen{L. Lindblom and G. Mendel,{\PRD{61}{104003}{2000}}}
\def\juergen{P. Papazoglou, J. Schaffner, S. Schramm, D. Zschiesche,
H. Stoecker and W. Greiner,{\PRC{55}{1499}{1997}}.}
\def\juergenn{P. Papazoglou, S. Schramm,  J. Schaffner-Bielich,
H. Stoecker and W. Greiner,{\PRC{57}{2576}{1998}}.}
\def\verena{V. Dexheimer, S. Schramm and D. Zschiesche, {\PRC{77}{025803}{2008}}}
\def\amruta{A. Mishra, A. Kumar, S. Sanyal, V. Dexheimer and S. Schramm,
arXiv:0905.3518[nucl-th].}
\def\juergennn{D. Zschiesche, J. Schaffner, L. Tolos, J. Schaffner-Bielich,
 R. Pisarski,{\PRC{75}{055202}{2007}}.}
\def\future {H. Mishra and V. Sreekanth, (in progress)}
\def\colsc{
 S.B. Ruester, I.A. Shovkovy and D.H. Rischke,{\JPG {31}{S849}{2005}};
K. Rajagopal and A. Schmitt, {\PRD{73}{045003}{2006}};
K. Rajagopal and R. Sharma, {\PRD{74}{094019}{2006}}; 
F. Neumann, M. Buballa and M. Oertel{\NPA{714}{481}{2003}}.}
\def\andersonn{N. Andersson, Astrophysics. J. {\bf 502},708 (1998); ibid,
Classical Quantum Gravity {\bf 20}, R105 (2003).}
\def\lindblom{L. Lindblom, arXiv:astro-ph/0101136.}
\def\prakash87{M. Prakash and T.L. Ainsworth,{\PRC{36}{346}{2987}}.}
\def\glendenningnpa{N.K. Glendenning{\NPA{480}{597}{1988}}.}
\def\pksahu{P.K. Sahu and A. Ohninshi, Prog. Theor. Phys. {\bf 104},1163(2000).}
\def\pksahuu{P.K. Sahu, T.K. Jha, K.C. Panda and S.K. patra,{\NPA{733}{169}{2004}}.}
\def\provost{J. Provost, G. berthomeiu and A. Rocca, Astron.
Astrophys. {\bf 94}, 126 (1981).}
\def\papaloizou{J. Papaloizou and J.E. Pringle, Mon. Not. R. Astron. Soc.
{\bf 182}, 423 (1978).}
\def\rischkehuang{Xu-Guang Huang, mei Huang, Dirk H Rischke and Armen
Sedrakian, arXiv:0910.3633v2[astro-ph].}

\end{document}